\documentclass[twocolumn,pre,showpacs,eqsecnum]{revtex4}

\usepackage{dcolumn}
\usepackage{amsfonts}
\usepackage{amsmath}
\usepackage{amssymb}
\usepackage{bm}
\usepackage{graphicx}
\usepackage{amssymb}

\newcommand{\brm}[1]{\bm{{\rm #1}}}
 
\begin{document}
\title{{\bf Corrections to Scaling in Random Resistor Networks and Diluted
Continuous Spin Models near the Percolation Threshold}}

\author{{\sc Hans-Karl Janssen}}
\affiliation{
Institut f\"{u}r Theoretische Physik III,
Heinrich-Heine-Universit\"{a}t,
40225 D\"{u}sseldorf,
Germany
}

\author{{\sc Olaf Stenull}}
\affiliation{
Department of Physics and Astronomy,
University of Pennsylvania,
Philadelphia, PA 19104,
USA
}
\vspace{10mm}
\date{\today}

\begin{abstract}
\noindent
We investigate corrections to scaling induced by irrelevant operators in randomly diluted systems near the percolation threshold. The specific systems that we consider are the random resistor network and a class of continuous spin systems, such as the $x$-$y$-model. We focus on a family of least irrelevant operators and determine the corrections to scaling that originate from this family. Our field theoretic analysis carefully takes into account, that irrelevant operators mix under renormalization. It turns out that long standing results on corrections to scaling are respectively incorrect (random resistor networks) or incomplete (continuous spin systems).  
\end{abstract}

\maketitle

\section{Introduction}
In the mid 1980's Harris, Lubensky and coworkers (HL)~\cite{harris_lubensky_84,harris_kim_lubensky_84,harris_lubensky_87b} developed a seminal field theoretic model for the unified description of percolating random resistor networks (RRNs) and a class of diluted continuous spin systems (such as the $x$-$y$-model). Their approach, based on ideas by Stephen~\cite{stephen_78}, turned out to be very fruitful. In the course of the years, it provided a foundation for the exploration of various critical properties, not only of the $x$-$y$-model and the RRN~\cite{harris_kim_lubensky_84,harris_lubensky_87b,stenull_janssen_oerding_99,stenull_janssen_oerding_2001}, but also of random resistor diode networks~\cite{janssen_stenull_prerapid_2001,stenull_janssen_jsp_2001}, the swiss cheese model~\cite{lubensky_tremblay_86_andmore,stenull_janssen_pre_2001_continuum} and random networks of Josephson junctions~\cite{john_lubensky_85,janssen_stenull_future}. Moreover, it fostered the computation of several fractal dimensions of isotropic~\cite{harris_87,janssen_stenull_oerding_99,janssen_stenull_99} and directed~\cite{janssen_stenull_prerapid_2001,stenull_janssen_pre_2001_nonlinear} percolation clusters and aided studying multifractality in isotropic~\cite{park_harris_lubensky_87,stenull_janssen_epl_2000,stenull_janssen_2001} and directed~\cite{stenull_janssen_epl_2001,stenull_janssen_pre_2002} percolation. Also, the HL model helped to improve the understanding of the vulcanization transition~\cite{janssen_stenull_pre_2001_vulcanization}.

An important role in the HL model is played by the exponent $\phi$. It describes the power law behavior of  the average resistance $M_{R}^{(1)}$ between two connected points $x$ and $x^{\prime }$ at the percolation threshold,
\begin{equation}
\label{ResScaling}
M_{R}^{(1)}\sim |x-x^{\prime }|^{\phi /\nu }\, .
\end{equation}
Here, $\nu $ is the critical exponent of the correlation length $\xi \sim |p-p_{c}|^{-\nu }$, where $p$ is the probability that controls the dilution of the network and $p=p_c$ marks the critical point. 

Originally~\cite{harris_kim_lubensky_84} $\phi$ was thought to be the first member, $\phi =\phi_{1}$, of an entire family $\{\phi _{m},m=1,2,\ldots \}$ of exponents with the $m$th member describing the $m$th cumulant of the resistance,
\begin{equation}
M_{R}^{(m)}\sim |x-x^{\prime }|^{\phi _{m}/\nu }\, .
\end{equation}
The $\phi_{m }$ were calculated to one-loop order~\cite{harris_kim_lubensky_84}. Shortly later Rammal, Lemieux, and Tremblay~\cite{rammal_lemieux_tremlay_85} showed that the higher crossover exponents  $\phi_{m > 1}$ are irrelevant and that higher cumulants scale as 
\begin{equation}
\label{MRscaling}
M_{R}^{(m)}\sim |x-x^{\prime }|^{m\phi /\nu } \, ,
\end{equation}
i.e., they display so called gap scaling. Furthermore, Rammal, Lemieux, and Tremblay argued that the $\phi_{m > 1}$ are important for corrections to scaling and specifically that these are governed by exponents $\omega _{m}=(m\phi -\phi _{m} )/\nu$. For example, the average resistance was thought to behave as
\begin{equation}
\label{wrong}
M_{R}^{(1)}\sim |x-x^{\prime }|^{\phi /\nu }\ \Big[1+\sum_{m }A_{m}\, |x-x^{\prime }|^{-\omega _{m}} \Big]\, ,
\end{equation}
and the conductivity $\Sigma$ as
\begin{equation}
\Sigma \sim (p-p_{c})^{t}\ \Big[1+\sum_{m}B_{m}\, (p-p_{c})^{\nu\omega _{m}}\Big] \, .
\end{equation}
Here, $t$ is the conductivity exponent $t=(d-2)\nu +\phi $, and $A_{m}$ and $B_{m}$ are non-universal amplitudes.

In this paper we take up the issue of corrections to scaling in RRN anew. Our careful analysis reveals that previous work on this subject is erroneous and that the $\phi_{m > 1}$ are meaningless, at least as far as corrections to scaling in RRN are concerned. It turns out that the previous studies overlooked a crucial feature of irrelevant field theoretic operators, viz.\ that they tend to mix under renormalization. Taking this subtlety into account, we calculate corrections to scaling for the average resistance and the conductivity. Moreover, we re-examine corrections to scaling in continuous spin models. Due to a difference in the symmetry properties that plays no role for the leading scaling behavior, the corrections to scaling in continuous spin models and RRN are somewhat different. We determine the correction to scaling exponents for the first cumulant of the spin orientations. 

The outline of our paper is as follows. In Sec.~\ref{HLmodel} we present a few basics about the RRN and the $x$-$y$-model. Then we state the field theoretic Hamiltonian defining the HL model and sketch its physical contents. Next we collect the irrelevant field theoretic operators that lead to the corrections to scaling we are interested in. These operators can be classified into 2 groups, viz.\ specific operators and general percolation operators. Section~\ref{review} briefly reviews known field theoretic results for the HL model including the leading scaling behavior and the leading correction to it. In Sec.~\ref{eom} we analyze the equation of motion implicit in the HL Hamiltonian. This analysis provides us with several results that are valid to arbitrary order in perturbation theory. Section~\ref{so} comprises our renormalization group analysis of the specific operators. The general percolation operators are scrutinized in Sec.~\ref{gpo}. In Sec.~\ref{scalingResults} we derive our final results for the critical behavior of the average resistance etc. The main part of our paper concludes with several remarks given in Sec.~\ref{concludingRemarks}. There are 3 Appendices. In Appendix~\ref{app:ward} we derive two useful identities that help us exploiting the equation of motion. Appendix~\ref{app:composite} features some general considerations on composite fields in the HL model. Appendix~\ref{app:XXX} contains details on the computation of Feynman diagrams.

\section{The Harris-Lubensky Model}
\label{HLmodel}

\subsection{Random resistor networks and the diluted $x$-$y$-model}
Both the RRN and the diluted $x$-$y$-model can be described by a Hamiltonian of the type 
\begin{equation}
\label{genHamil}
H = - \sum_{\langle i,j \rangle} U_{i,j} \left( \vartheta_i - \vartheta_j \right) \, ,
\end{equation} 
where $\vartheta_i$ is a continuous dynamical variable pertaining to a site $i$ of a $d$-dimensional hypercubic lattice and the sum runs over all nearest neighbor pairs on this lattice. For the RRN, $\vartheta_i$ corresponds to the voltage $V_i$ at site $i$ and is defined on the interval $[-\infty , \infty]$. 
\begin{equation}
U_{i,j} \left( V\right) = - \frac{1}{2} \, \sigma_{i,j} V^2 
\end{equation}
is the electrical power dissipated on the bond between $i$ and $j$, with $\sigma_{i,j}$ denoting the conductance of this bond. $\sigma_{i,j}$ is a random variable that takes on the values $\sigma$ and $0$ with probability $p$ and $1-p$, respectively. In the case of the $x$-$y$-model, $\vartheta_i$ is the angle $\varphi_i$ that specifies the orientation of the spin at site $i$ and is defined on the interval $[-\pi , \pi]$. Here, 
\begin{equation}
U_{i,j} \left( \varphi \right) = K_{i,j} \cos \varphi \,  
\end{equation}
with $K_{i,j}$ being the exchange integral. In the diluted $x$-$y$-model the exchange integral is assumed to take on the values $K$ and $0$ respectively with probability $p$ and $1-p$.

\subsection{The Harris-Lubensky Hamiltonian}
Based on the Hamiltonian~(\ref{genHamil}) HL derived a field theoretic model that can be written as
\begin{eqnarray}
{\cal H} &=&\int d^{d}x\,\biggl\{\sum_{\vec{\theta}}\biggl(\frac{\tau }{2}
s^{2}+\frac{1}{2}(\nabla s)^{2}+\frac{w}{2}(\nabla _{\theta }s)^{2}+\frac{g}{
6}s^{3}\biggr)
\nonumber \\
&+& \sum_{i}f_{i}{\cal A}_{i}\biggr\}\, .  \label{Hamilt}
\end{eqnarray}
For details on the derivation we refer the reader to Ref.~\cite{harris_lubensky_87b}. The order parameter field $s({\bf x},\vec{\theta})$ lives on a continuous $d$-dimensional space with the coordinates ${\bf x}$. It is subject to the constraint
\begin{equation}
\label{constraint}
\sum_{\vec{\theta}}s({\bf x},\vec{\theta})=0  \, .
\end{equation}
The variable $\vec{\theta}$ is a replicated analog of the original dynamical variable $\vartheta$ and takes on discrete values on a $D$-dimensional torus (the replica space). Formally, $\vec{\theta}$ is given by $\vec{\theta}=\vec{k}\Delta \theta$, where $\vec{k}$ is a $D$-dimensional vector with integer components $k^{(\alpha )}$ and $-M<k^{(\alpha )}\leq M$. To recover the physical situation, on has to take the replica limit $D\rightarrow 0$, $M\rightarrow \infty$ with $M^{D}\rightarrow 1$ and $\Delta \theta =\theta _{0}/\sqrt{M}$. In this limit, $\theta _{0}$ plays the role of a redundant scaling
parameter, i.e., the theory is independent of its value. The parameter $\tau$ is proportional to $p_c -p$, i.e., it specifies the distance from the critical point. $w$ is proportional to $\sigma^{-1}$ or $K^{-1}$, respectively. The ${\cal A}_{i}$, finally, are irrelevant field theoretic operators (rotationally invariant monomials constructed from the fundamental field $s$ and its derivatives in real and replica space). Specifics of the irrelevant operators will be given below.

For $w=f_i=0$, the HL Hamiltonian describes the $N$-state Potts model with $N=\sum_{\vec{\theta}}1=(2M)^{D}$. In this case we have $S_{N}$, the group of all permutations of $N$ objects, as the internal symmetry group. If $w\neq 0$, this symmetry is reduced to $O(D)$, the group of orthogonal rotations in the replica space. A particular scaling symmetry of the Hamiltonian will be important as we go along. Namely, ${\cal H}$ is invariant under the rescaling $w \to b \, w$  because the scaling factor $b$ can be absorbed into the redundant parameter $\theta_0$.

In the following we use that $\sum_{\vec{\theta}}\ldots \approx (\Delta \theta )^{D}\sum_{\vec{
\theta}}\ldots \approx \int d^{D}\theta \ldots$ and abbreviate the latter integral by $\int_{\vec{\theta}}\ldots $. The approximations involved here become exact in the replica limit.

\subsection{Physical contents}
To fully appreciate the physical contents of the HL model it is helpful to consider the replica space Fourier transform
\begin{equation}
\label{FourierT}
\psi_{\vec{\lambda}} (\brm{x}) =  \int_{\vec{\theta}} \exp \left( - i\vec{\lambda} \cdot \vec{\theta} \right)  s({\bf x},\vec{\theta})  
\end{equation}
of the order parameter. For completeness we mention here that the constraint~(\ref{constraint}) translates upon replica space Fourier transform into
\begin{equation}
\label{constraint2}
\psi_{\vec{0}} (\brm{x}) =  0 \, .
\end{equation}
This constraint is intuitively clear because $\psi_{\vec{0}} (\brm{x})$ is a constant and hence does no qualify as an order parameter. It will play an important role in our renormalization group (RG) analysis.  

The value of the quantity $\psi_{\vec{\lambda}} (\brm{x})$ is that its correlation functions
\begin{equation}
\label{corrPsi}
G \left( \brm{x}, \brm{x}^\prime ; \vec{\lambda} \right) = \left\langle  \psi_{\vec{\lambda}} (\brm{x}) \psi_{- \vec{\lambda}} (\brm{x}^\prime )\right\rangle
\end{equation}
provide convenient access to physical observables. In the case of the RRN one has
\begin{equation}
\label{corrPsiRRN}
G \left( \brm{x}, \brm{x}^\prime ; \vec{\lambda} \right) = \left\langle \exp \left[ - \frac{\vec{\lambda}^2}{2} \, R (\brm{x}, \brm{x}^\prime) \right] \right\rangle_C \, ,
\end{equation}
where $R (\brm{x}, \brm{x}^\prime) = \langle [V_{\brm{x}} - V_{\brm{x}^\prime}]^2\rangle$ is the macroscopic resistance between the points $\brm{x}$ and $\brm{x}^\prime$ and $\langle \cdots \rangle_C$ denotes averaging over all configurations of the diluted lattice. Hence, $G ( \brm{x}, \brm{x}^\prime ; \vec{\lambda} )$ is a generating function for the moments of the resistance distribution
\begin{equation}
\label{defMR}
M_R^{(l)}  (\brm{x}, \brm{x}^\prime) = \left\langle R (\brm{x}, \brm{x}^\prime)^l   \right\rangle_C^\prime \, ,
\end{equation}
where the prime indicates averaging subject to the constraint that $\brm{x}$ and $\brm{x}^\prime$ are connected. 

For the diluted $x$-$y$-model the Hamiltonian~(\ref{genHamil}) is not Gaussian and one is led to
\begin{eqnarray}
\label{corrPsiXY}
&&G \left( \brm{x}, \brm{x}^\prime ; \vec{\lambda} \right) = 
 \\
&& \left\langle \exp \left[ \sum_{l=1}^\infty \frac{(-1)^l}{(2l)!} \, K_l (\vec{\lambda}) \left\langle \left[ \varphi_{\brm{x}} -\varphi_{\brm{x}^\prime} \right]^{2l} \right\rangle^{(c)} \right] \right\rangle_C  ,
\nonumber
\end{eqnarray}
where $\langle \cdots \rangle^{(c)}$ stands for the cumulants with respect to the average $\langle \cdots \rangle$ and $K_l (\vec{\lambda}) = \sum_{\alpha =1}^D [\lambda^{(\alpha)}]^{2l}$. Thus, $G ( \brm{x}, \brm{x}^\prime ; \vec{\lambda} )$ is a generating function for the cumulants
\begin{equation}
\label{defCphi}
C_\varphi^{(l)}  (\brm{x}, \brm{x}^\prime) = \left\langle \left\langle \left[ \varphi_{\brm{x}} -\varphi_{\brm{x}^\prime} \right]^{2l} \right\rangle^{(c)}  \right\rangle_C^\prime \, ,
\end{equation}
which measure the fluctuation of the angular variables. $C_\varphi^{(1)}$, in particular, is related to the spin-wave stiffness. 

\subsection{The  irrelevant operators ${\cal A}_{i}$}
The main goal of this paper is to analyze corrections to scaling. The leading correction to scaling for the average resistance etc.\ is well known since it is described by the so-called Wegner exponent, see Sec.~\ref{leader} below. The next to leading corrections stem from irrelevant operators which scale as ${\cal A}_{i}\sim \mu ^{8}$ at the upper critical dimension $d_{c}=6$, where $\mu$ is some inverse length scale. These operators can be classified into two groups. The first group consist of operators having at least two derivatives with respect to the replica space. The operators belonging to this group will be referred to as specific operators. In the case of the RRN, the specific operators are given by the $O(D)$ invariant composite fields
\begin{subequations}
\label{ops1}
\begin{eqnarray}
{\cal A}_{0} &=&\frac{w}{2}\nabla ^{2}\Big(\int_{\vec{\theta}}s\nabla _{\theta}^{2}s\Big)\, , \\
{\cal A}_{1} &=& \frac{w^{2}}{2}\int_{\vec{\theta}}\bigl(\nabla_{\theta }^{2}s\bigr)^{2} \, , \\
{\cal A}_{2} &=&\frac{w}{2}\int_{\vec{\theta}}s\nabla ^{2}\nabla _{\theta }^{2}s \, , \\
{\cal A}_{3} &=& -\frac{w}{6}\int_{\vec{\theta}}s^{2}\nabla _{\theta }^{2}s \, .
\end{eqnarray}
\end{subequations}
Each replica space derivative is accompanied by a factor $\sqrt{w}$ to ensure invariance under the rescaling $w \to b \, w$. 

When considering the $x$-$y$-model, we have to admit all of the hypercubic invariants of $\nabla_{\theta}$ rather than of $\nabla_{\theta}^2$. Thus, we are confronted with a further operator with the naive dimension 8, namely  
\begin{equation}
\label{Ac}
{\cal A}_{c}=\frac{w^{2}}{2}\int_{\vec{\theta}}\sum_{\alpha =1}^{D}\bigl(\nabla_{\theta^{(\alpha) }}^{2}s\bigr)^{2}\, .
\end{equation}
Note that this operator breaks the $O(D)$-symmetry in replica space.

As we go along we will see that the family $\{{\cal A}_{0}, \cdots, {\cal A}_{3} \}$ is associated with corrections to scaling of the average resistance and conductivity in RRNs. Hence, we refer to this family as resistor specific. The larger family $\{{\cal A}_{c}, {\cal A}_{0}, \cdots, {\cal A}_{3} \}$ is associated with corrections of the spin orientation cumulants in diluted continuous spin systems. Thus, we say this family is spin specific. 

The next group of operators leads to corrections for both pure percolation
and the respective specific behavior. We refer to these operators as general percolation operators. The pure percolation behavior results from the HL model in the limit $w=0$. Thus, the second group involves only operators without derivatives with respect to the replica space. They are given by the $S_{N}$ invariant composite fields 
\begin{subequations}
\label{ops3}
\begin{eqnarray}
{\cal A}_{4} &=&\frac{1}{2}\int_{\vec{\theta}}\bigl(\nabla ^{2}s\bigr)^{2}\, , \\
{\cal A}_{5} &=&-\frac{1}{6}\int_{\vec{\theta}}s^{2}\nabla ^{2}s\, , \\
{\cal A}_{6} &=&\frac{1}{4!}\Big(\int_{\vec{\theta}}s^{2}\Big)^{2}\, , \\
{\cal A}_{7} &=&\frac{1}{4!}\int_{\vec{\theta}}s^{4}\, .
\end{eqnarray}
\end{subequations}
At first sight, three additional general percolation operators with naive dimension eight seem to matter, namely
\begin{subequations}
\label{ops4}
\begin{eqnarray}
\mathcal{A}_{8}  &=&\frac{1}{2} \int_{\vec{\theta}}s \bigl(\nabla^{2}\bigr)^{2}s\,,
\\
\mathcal{A}_{9}  &=&\frac{1}{2}\nabla^{2}\int_{\vec{\theta}}s\nabla ^{2}s\,,
\\
\mathcal{A}_{10}  &=&\frac{1}{2}\bigl(\nabla^{2}\bigr)^{2}\int_{\vec{\theta}}s^{2}\,,
\\
\mathcal{A}_{11}  & =&\frac{1}{6}\nabla^{2}\int_{\vec{\theta}}s^{3}\, . 
\end{eqnarray}
\end{subequations}
However, these operators can be neglected in calculating corrections to scaling. Upon Fourier transformation of ${\cal A}_{4}$ and $\mathcal{A}_{8}$ one sees that these operators coincide for vanishing external momentum. Thus, it is sufficient foe our purposes to keep ${\cal A}_{4}$. $\mathcal{A}_{9}$ to $\mathcal{A}_{11}$ are merely total derivatives of lower dimensional operators. Their Fourier transformed counterparts are proportional to some (positive and even) power of a external momentum and hence these operators cannot contribute to a translational invariant Hamiltonian. This situation is similar for $\mathcal{A}_{0}$. We choose, though, to keep $\mathcal{A}_{0}$ in our analysis for 2 reasons: (i) To exemplify explicitly that this kind of operator does in the end not contribute to the corrections to scaling. (ii) Based on the scaling symmetries of $\mathcal{H}$ we can draw exact conclusions on $\mathcal{A}_{0}$ which then can be compared to the results of our explicit 1-loop calculation, i.e., retaining $\mathcal{A}_{0}$ allows for a later consistency check.

\section{A brief review of known renormalization group results}
\label{review}
Many of the critical properties of the HL model are well known. Substantial contributions stem, inter alia, from HL. In earlier work we have investigated the HL by using the powerful methods of renormalized field theory. Here we briefly review parts of this work to provide background and to establish notation.

\subsection{Renormalization and Scaling}
In Ref.~\cite{stenull_janssen_oerding_99} we have studied the HL model with all the $f_{i}$ equal to zero. In particular we have determined the renormalizations
\begin{subequations}
\label{reno} 
\begin{align}
s \rightarrow \mathaccent"7017{s}&=Z^{1/2}s\, ,  \\
\tau \rightarrow \mathaccent"7017{\tau }&=Z^{-1}Z_{\tau }\tau \, ,\\
w \rightarrow \mathaccent"7017{w}&=Z^{-1}Z_{w}w\, ,   \\
g \rightarrow \mathaccent"7017{g}&= Z^{-3/2}Z_{g}g\, , \\
u & = G_{\varepsilon } \mu^{-\varepsilon} g^2
\end{align}
\end{subequations}
in dimensional regularization and minimal subtraction of $\varepsilon$-poles~\cite{amit_zinn-justin}. Here, the $\mathaccent"7017{}$ denotes bare, unrenormalized quantities. The coupling constant $u$ is introduced because it is convenient and dimensionless.  $\varepsilon =6-d$ measures the deviation from the upper critical dimension. $G_{\varepsilon }=\Gamma (1+\varepsilon /2)/(4\pi )^{d/2}$ is a dimension-dependent numerical factor. We have determined the renormalization constants $Z_{\cdots }(u)$ to second order in $u$ in a 2-loop calculation. From the scheme~(\ref{reno}) we derived a Gell-Mann-Low RG equation for the vertex functions $\Gamma^{(n)}$ with $n$ amputated external legs,
\begin{eqnarray}
\label{rge}
&&\left[ \mu \frac{\partial }{\partial \mu} + \beta \frac{\partial }{\partial u} + \tau \kappa \frac{\partial }{\partial \tau} + w \zeta \frac{\partial }{\partial w} - \frac{n}{2} \gamma \right]  
\nonumber \\
&& \times \, \Gamma^{(n)} \left( \left\{ {\rm{\bf x}} ,\vec{\lambda} \right\} ; \tau, u, w, \mu \right) = 0 \, ,
\end{eqnarray}
with the Wilson functions $\beta \left( u \right) = \mu \frac{\partial u}{\partial \mu} |_0$, $\kappa \left( u \right) = \mu \frac{\partial \ln \tau}{\partial \mu}  |_0$, $\zeta \left( u \right) = \mu \frac{\partial \ln w}{\partial \mu} |_0$ and $\gamma \left( u \right) = \mu \frac{\partial }{\partial \mu} \ln Z |_0$~\cite{footnote1}. Solving Eq.~(\ref{rge}) at the infrared stable fixed point $u^{\ast }$ leads to the scaling behavior of the vertex functions~\cite{footnote2},
\begin{eqnarray}
\label{scaling}
&&\Gamma^{(n)} \left( \left\{ {\rm{\bf x}} ,\vec{\lambda} \right\} ; \tau, u, w, \mu \right) = 
\ell^{-(d-2+\eta)n/2} 
\nonumber \\
&& \times \,
\Gamma^{(n)} \left( \left\{ \ell{\rm{\bf x}} , \vec{\lambda} \right\} ; \ell^{-1/\nu}\tau , u^\ast,  \ell^{-\phi /\nu}w, \mu \right) \, .
\end{eqnarray}
The exponents $\eta = \gamma^\ast$, $\nu = (2-\kappa^\ast)$, and $\phi = \nu (2-\zeta^\ast)$ [$\gamma^\ast= \gamma (u^\ast)$, $\kappa^\ast= \kappa (u^\ast)$ and so on] are the usual critical exponents for percolation and the RRN. In the present paper we work to 1-loop order. To this order, $u^\ast = 2\varepsilon/7+ O ( \varepsilon^2 )$ and the critical exponents are given by $\eta = - \varepsilon/21 + O ( \varepsilon^2 )$, $\nu = 1/2 + 5\varepsilon/84+ O ( \varepsilon^2 )$ and $\phi = 1+ \varepsilon/42+ O ( \varepsilon^2 )$.

From (\ref{scaling}) one can extract the leading scaling behavior of various observables. Exploiting Eq.~(\ref{corrPsiRRN}), for example, it is straightforward to derive Eq.~(\ref{MRscaling}) for  $M_{R}^{(m)}$.

\subsection{The leading correction to scaling}
\label{leader}
The leading correction to scaling is, as usual, governed by the so-called Wegner exponent. This leading correction emerges when the renormalized coupling $u$ is not exactly equal $u^\ast$ since the renormalization flow has not arrived at its fixed point yet. Such a case occurs typically when there is a finite momentum cutoff reminiscent of a non-vanishing lattice spacing.

Taking the leading correction into account, the scaling behavior, e.g.,  of the average resistance~(\ref{ResScaling}) becomes 
\begin{equation}
\label{wegner}
M_{R}^{(1)}\sim |x-x^{\prime }|^{\phi /\nu }\ \big[1+A |x-x^{\prime }|^{-\omega }\big]\, ,
\end{equation}
where $A$ is a non-universal amplitude and  $\omega = \beta^\prime \left( u^\ast \right)$ is the Wegner exponent. $\omega$ can be calculated without much effort to third order in $\varepsilon$ upon using the three-loop result for $\beta \left( u \right)$ obtained by de Alcantara Bonfim {\it et al}.~\cite{alcantara_80}. To the order we are working here, the Wegner exponent is given by
\begin{eqnarray}
\omega = \varepsilon + O \left( \varepsilon^2  \right) \, .
\end{eqnarray}
Expression similar to (\ref{wegner}) hold for the conductivity and so on. 

\section{Consequences of the Equation of Motion}
\label{eom}
Several of the correction to scaling exponents originating from the specific and the general percolation operators can be derived without resorting to an explicit perturbation calculation. We will do so by analyzing the classical equation of motion. The so obtained results have the virtue of being rigorous in the sense that they hold to arbitrary in perturbation theory. In the remainder we will frequently encounter the so-called scaling dimension $x_{{\cal A}}$ of an eigenoperator ${\cal A}$. Formally, $x_{{\cal A}}$ is defined via the rescaling ${\cal A}({\bf x})=\ell^{x_{{\cal A}}}{\cal A}(\ell{\bf x})$ when ${\cal A}$ is inserted into vertex or correlation functions at the critical point.

Before we turn to the consequences of the equation of motion we take a little detour and extract a consequence of the invariance of $\mathcal{H}$ under the rescaling $w \to b \, w$. It follows from this rescaling invariance that, in the case $w\neq 0$, the operator $w\int_{\vec{\theta}}(\nabla _{\theta }s)^{2}$ is marginal and its scaling dimension is equal to $d$. Note that ${\cal A}_{0}$ results from the application of the Laplacian $\nabla ^{2}$ to the this marginal operator. Consequently the scaling dimension of ${\cal A}_{0}$ is given by 
\begin{eqnarray}
x_{{\cal A}_{0}}=d+2\, . 
\end{eqnarray}

Now we turn to the equation of motion. It is well known that for every independent equation, that follows from the equation of motion, there is an eigenoperator of the RG with a scaling dimension that can be expressed in terms of the scaling dimensions of lower-dimensional operators. For some background on four- and six-dimensional operators in the HL model we refer to Appendix~\ref{app:composite}.

The classical equation of motion derived from the Hamiltonian (\ref{Hamilt}) and the constraint (\ref{constraint}) reads
\begin{eqnarray}
&&{\cal H}^{\prime }:=\left. \frac{\delta {\cal H}}{\delta s}\right|_{f=0}
\nonumber \\
&&=\, -\nabla ^{2}s-w\nabla _{\theta }^{2}s+\tau s+\frac{g}{2}\Big(s^{2}-
\frac{1}{N}\int_{\vec{\theta}}s^{2}\Big)=0\, . \qquad
\end{eqnarray}
In the following we work in the limit $D\rightarrow 0$, i.e., $N=(2M)^{D}\rightarrow 1$. We consider the lower dimensional operators
\begin{equation}
{\cal B}_{1}=w\nabla _{\theta }^{2}s\ ,\qquad {\cal B}_{2}=\nabla ^{2}s\
,\qquad {\cal B}_{3}=\mathcal{H}^{\prime}\, .
\end{equation}
The scaling dimensions of the first two follow from their renormalizations as
\begin{equation}
x_{{\cal B}_{1}}=\frac{d-2+\eta }{2}+\frac{\phi }{\nu }\, ,\qquad x_{{\cal B}
_{2}}=\frac{d+2+\eta }{2}\, .
\end{equation}
The scaling dimension of ${\cal B}_{3}$,
\begin{equation}
x_{{\cal B}_{3}}=d-2+\eta _{{\cal B}_{3}}\, ,
\end{equation}
 is not the same as the scaling dimension of the operator ${\cal B}_{0}=\int_{\theta }s^{2}$. ${\cal B}_{0}$ belongs to the trivial representation of the permutation symmetry group $S_{N}$, whereas the operators ${\cal B}_{i}$ with $i=1, 2,3$ transform like $s({\bf x},\vec{\theta})$, i.e., they belong  to the fundamental representation of $S_{N}$ [permutations of $N$ objects $\phi_{i}$ with constraint $\sum_{i=1}^{N}\phi_{i}=0$]. Hence, $\eta _{{\cal B}_{3}}$ remains to be determined. In Appendix~\ref{app:ward} we derive a Ward identity, Eq.~(\ref{wardHPrime}), that implies
\begin{equation}
\eta _{{\cal B}_{3}}=\frac{\varepsilon - \eta}{2}
\end{equation}
to arbitrary order in $\varepsilon$ expansion. Consequently, we obtain the exact result
\begin{equation}
x_{{\cal B}_{3}}=\frac{d+2-\eta }{2} \, .
\end{equation}

Now we can extract the actual consequences of the equation of motion for our prime targets of interest, viz.\ the eight-dimensional operators. In Appendix~\ref{app:ward} we show that
\begin{eqnarray}
\label{anIdentity}
& & \left\langle s({\bf x}_{1},\vec{\theta}_1)\cdots s({\bf x}
_{n},\vec{\theta}_n) \, \mathcal{F} \cdot \mathcal{H}^{\prime} (\brm{x}) \right\rangle =\sum_{i=1}^{n}\delta ({\bf x}-{\bf x}_{i})
\nonumber \\
& & \times \,  \Big\langle s({\bf x}_{1},\vec{\theta}_1)\cdots s({\bf x}_{i-1},\vec{\theta}_{i-1})\, \mathcal{F} ({\bf x}_{i},\vec{\theta}_i) 
\nonumber \\
&&\times \,  s({\bf x}_{i+1},\vec{\theta}_{i+1})\cdots s({\bf x}_{n},\vec{\theta}_{n})\Big\rangle \,  . \quad
\end{eqnarray}
where $\mathcal{F}$ is some composite field and where we defined
\begin{equation}
\mathcal{F} \cdot \mathcal{H}^{\prime} (\brm{x}) = \int_{\vec{\theta}} \mathcal{F} ({\bf x},\vec{\theta}) \, \mathcal{H}^\prime ({\bf x},\vec{\theta}) \, .
\end{equation}
Upon renormalization, on finds that the identity~(\ref{anIdentity}) implies the scaling relation 
\begin{equation}
\label{scalingRel}
x_{\mathcal{F} \cdot \mathcal{H}^{\prime}}=d-x_{s}+x_{\mathcal{F}}=\frac{d+2-\eta }{2}+x_{\mathcal{F}}\, ,
\end{equation}
where we used that $x_s = (d-2+\eta)/2$. Specifying $\mathcal{F}$ as ${\cal B}_k$, $k=1, 2,3$, we obtain the (as yet unrenormalized) operators 
\begin{subequations}
\label{Eq.Mot.}
\begin{align}
{\cal B}_{1} \cdot {\cal H}^{\prime }&=-2({\cal A}_{1}+
{\cal A}_{2})-3g{\cal A}_{3}\, ,
 \\
{\cal B}_{2} \cdot  {\cal H}^{\prime }&=-2({\cal A}_{2}+
{\cal A}_{4})-3g{\cal A}_{5}\, , 
 \\
{\cal B}_{3} \cdot {\cal H}^{\prime }&=2(\mathcal{A}_{1}+2\mathcal{A}_{2}+ \mathcal{A}_{4}) \nonumber \\ & + 6g(\mathcal{A}_{3}+\mathcal{A}_{5}) -6g^{2}(\mathcal{A}_{6}-\mathcal{A}_{7})\, . 
\end{align}
\end{subequations}
These combinations are eigenoperators of the RG, at least at 0-loop order. At higher loop orders, the renormalization might modify the combinations appearing on the right hand sides of Eqs.~(\ref{Eq.Mot.}). Our explicit calculations presented in Secs.~\ref{so} and \ref{gpo} reveals however, that this effect is absent at 1-loop order. More importantly, general RG arguments (cf. Ref.~\cite{amit_zinn-justin}) guarantee that this effect has no impact on the scaling dimensions at any loop order. From Eq.~(\ref{scalingRel}) we readily deduce that 
\begin{subequations}
\label{eomScaleRel}
\begin{eqnarray}
x_{{\cal B}_{1} \cdot {\cal H}^{\prime }} &=&d+\frac{\phi }{\nu }\ , \\
x_{{\cal B}_{2} \cdot {\cal H}^{\prime }} &=& d+2\, ,  \\
x_{{\cal B}_{3} \cdot {\cal H}^{\prime }} &=& d+2-\eta \, .
\end{eqnarray}
\end{subequations}
Two points are worth being emphasized: (i) the scaling relations~(\ref{eomScaleRel}) are correct to arbitrary order in perturbation theory. (ii) Equations~(\ref{Eq.Mot.}) and (\ref{eomScaleRel}) allow us consistency checks of our explicit diagrammatic calculations. This is particularly valuable because these calculations involve fairly many diagrams and one is confronted with a certain risk of algebraic errors or erroneous symmetry factors. 

To prepare the ground for the announced consistency checks we now evaluate Eqs.~(\ref{eomScaleRel}) to 1-loop order. Exploiting the known $\varepsilon$ expansion results for $\phi$ and $\nu$ we find that the scaling dimensions of the eigenoperators $\mathcal{B}_{1} \cdot \mathcal{H}^{\prime}$ and $\mathcal{B}_{3} \cdot \mathcal{H}^{\prime}$ are given to the order we are working here by 
\begin{subequations}
\begin{align}
x_{{\cal B}_{1} \cdot {\cal H}^{\prime }}& = d+2-\frac{4\varepsilon }{21} +O(\varepsilon^{2}) \, ,
\\
\label{nice}
x_{{\cal B}_{3} \cdot {\cal H}^{\prime }}& = d+2+\frac{\varepsilon }{21}+O(\varepsilon ^{2})\, .
\end{align}
\end{subequations}
It is a nice little exercise to derive the result~(\ref{nice}) explicitly in a 1-loop calculation. An insertion of  ${\cal B}_{3}$ into the two-point vertex function $\Gamma ^{(2)}$
leads to 1-loop order to the singular part 
\begin{equation}
\mathaccent"7017{\Gamma }_{\mathaccent"7017{{\cal B}}
_{3}}^{(2)}(\{{\bf p}=0\})=-1+\frac{2u}{\varepsilon }\, .
\end{equation}
of the bare function. Thus, it arises a factor $2$ in the singular part in contrast to a
corresponding insertion of ${\cal B}_{0}$. Renormalizing $\mathaccent"7017{\cal B}_{3}=Z_{{\cal B}_{3}}{\cal B}_{3}$ we find, using $Z=1+\frac{u}{6\varepsilon}$, that $Z_{{\cal B}_{3}}=1-\frac{11u}{6\varepsilon }$.  From this result we eventually get 
\begin{equation}
\eta _{{\cal B}_{3}}=\frac{11\varepsilon }{21}+O(\varepsilon ^{2})\, ,
\end{equation}
which then leads to Eq.~(\ref{nice}).

\section{Renormalization of specific Operators}
\label{so}

\subsection{RRN specific operators}
\begin {figure}[ptb]
\includegraphics [width=6cm]{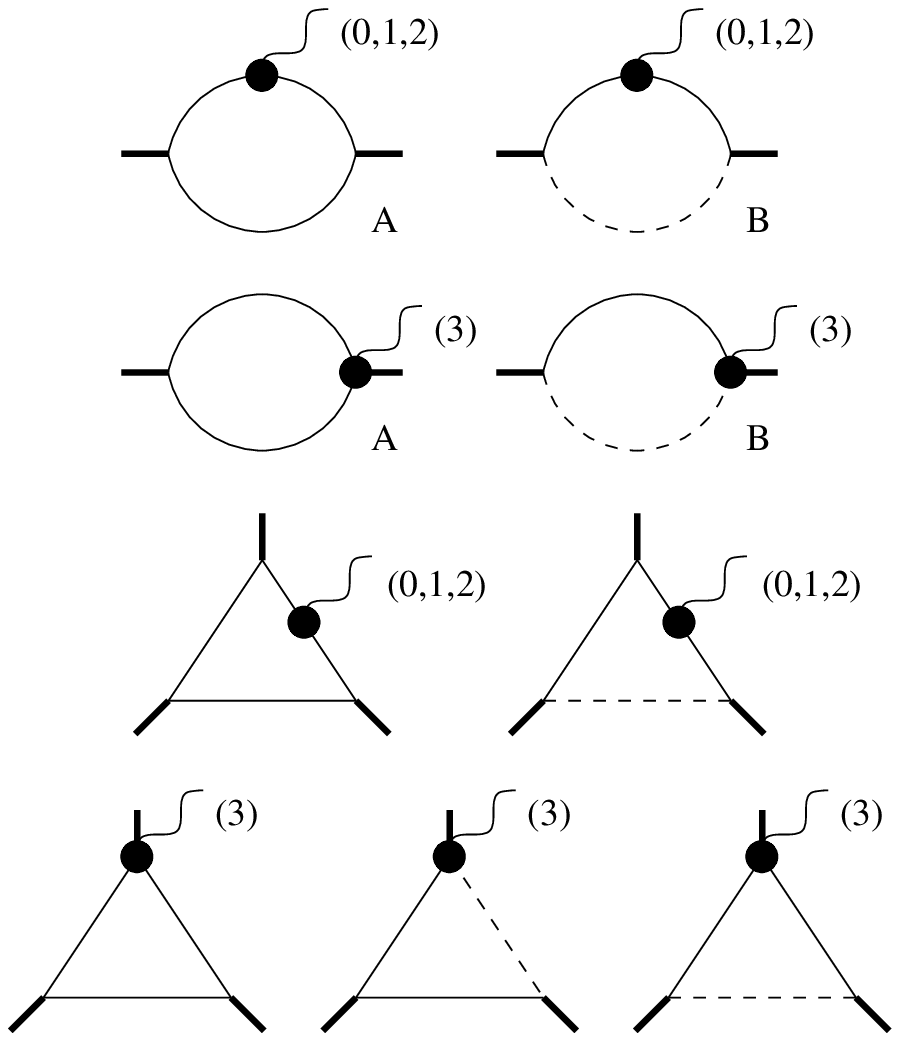}\caption {Diagrams contributing at 1-loop order to the renormalization of the RRN specific operators. These diagrams are obtained by first decomposing the bold 1-loop diagrams into their conducting diagrams and by then inserting the irrelevant operators in the appropriate places. The bold lines symbolize bold propagators, the light lines stand for conducting and the dashed lines for insulating propagators. The labels A and B refer to conducting diagrams before insertions are made and will be used in the appendix. The dots with wiggly lines stand for an insertion of (0) ${\cal A}_0^\prime$, (1) ${\cal A}_1^\prime$, (2) ${\cal A}_2^\prime$ and (3) ${\cal A}_3^\prime$, respectively.}
\label {fig1}
\end {figure}

To determine the renormalizations of the operators ${\cal A}_{0},\ldots ,{\cal A}_{3}$ we study their insertions into the vertex functions $\Gamma^{(n)}$.  We proceed in the spirit of our previous work on the RRN. We use 
\begin{equation}
G({\bf p},\vec{\lambda})=\langle \tilde{\psi}_{\vec{\lambda}}({\bf p})\, \tilde{\psi}_{-\vec{\lambda}} (-{\bf p})\rangle ^{(\text{trunc})}
\end{equation}
as the Gaussian propagator, where $\tilde{\psi}_{\vec{\lambda}}({\bf p})$ stands for the Fourier transform defined via
\begin{equation}
\tilde{\psi}_{\vec{\lambda}}({\bf x}) =\int_{{\bf p}}{\rm e}^{i{\bf p\cdot x}}\, \tilde{\psi}_{\vec{\lambda}}({\bf p})\, .
\end{equation}
Here, $\int_{{\bf p}}$ is an abbreviation for $(2\pi)^{-d/2} \int d^d p$. $\langle \cdots \rangle ^{(\text{trunc})}$ denotes averaging with respect to the Gaussian part of $\mathcal{H}$ and all $f_i =0$. Explicitly, our propagator reads
\begin{equation}
\label{propdeco}
G({\bf p},\vec{\lambda})=\frac{1-\delta _{\vec{\lambda},0}}{{\bf p}^{2}+w
\vec{\lambda}^{2}+\tau }=\frac{1}{{\bf p}^{2}+w\vec{\lambda}^{2}+\tau }-
\frac{\delta _{\vec{\lambda},0}}{{\bf p}^{2}+\tau } \, .
\end{equation}
Note that on the right hand side of Eq.~(\ref{propdeco}) the propagator is decomposed into two parts: one with unrestricted values of $\vec{\lambda}$ (conducting) and one with $\vec{\lambda}=0$ (insulating).
The decomposition of the propagator leads to a decomposition of each of the original (bold) Feynman diagrams into an assembly of diagrams made of conducting and insulating propagators. These conducting diagrams resemble essential features of real resistor networks. One can say that they have a real-world interpretation~\cite{stenull_janssen_oerding_99,stenull_janssen_oerding_2001,janssen_stenull_prerapid_2001,stenull_janssen_jsp_2001,stenull_janssen_pre_2001_continuum,janssen_stenull_future,janssen_stenull_pre_2001_vulcanization,janssen_stenull_oerding_99,janssen_stenull_99,stenull_janssen_pre_2001_nonlinear,stenull_janssen_epl_2000,stenull_janssen_2001,stenull_janssen_epl_2001,stenull_janssen_pre_2002}.  Upon recasting the conducting and insulating propagators in the Schwinger representation and using the continuum limit in the replica space with $D\rightarrow 0$, it is easy to calculate all required diagrams, see Fig.~\ref{fig1}, in dimensional regularization, at least to one-loop order.  For convenience, we work with the rescaled operators
\begin{equation}
\label{rescaledOps}
{\cal A}_{i }^{\prime }=\biggl(\frac{\mu ^{\varepsilon }}{
G_{\varepsilon }}\biggr)^{k_{i }/2}{\cal A}_{i }
\end{equation}
instead of the original ${\cal A}_{i }$. The $k_i$ are  $k_{0}=k_{1}=k_{2}=0$ and $k_{3}=1$. The benefit of this rescaling is that the primed operators all have the same naive dimension, viz.\ ${\cal A}_{i }^{\prime } \sim \mu ^{d+2}$.  Our 1-loop calculation sketched in Appendix~\ref{app:XXX} gives for  the singular parts of the primitively divergent vertex functions with operator insertions $\Gamma _{{\cal A}_{i }^{\prime }}^{(n)}$, $n=2$ and $3$, the results
\begin{subequations}
\label{resGamma2}
\begin{eqnarray}
-\Gamma _{{\cal A}_{0}^{\prime }}^{(2)}({\bf p}_{1},{\bf p}_{2},\vec{\lambda}
) &=&w\vec{\lambda}^{2}{\bf p}^{2}\biggl(1-\frac{5u}{6\varepsilon }\biggr)\, ,
 \\
-\Gamma _{{\cal A}_{1}^{\prime }}^{(2)}({\bf p}_{1},{\bf p}_{2},\vec{\lambda}
) &=&w\vec{\lambda}^{2}\bigg[ w\vec{\lambda}^{2}\biggl(1-\frac{u}{
\varepsilon }\biggr)
\nonumber \\
&-& \frac{u}{\varepsilon }\biggl(\frac{{\bf p}^{2}}{90}+
\frac{{\bf p}_{1}^{2}{\bf +p}_{2}^{2}}{30}\biggr) \bigg] \, ,  
 \\
-\Gamma _{{\cal A}_{2}^{\prime }}^{(2)}({\bf p}_{1},{\bf p}_{2},\vec{\lambda}
) &=&w\vec{\lambda}^{2}\bigg[ \frac{{\bf p}_{1}^{2}{\bf +p}_{2}^{2}}{2}
\biggl(1-\frac{3u}{10\varepsilon }\biggr)
\nonumber \\
&+& \frac{u}{\varepsilon }\biggl(\frac{
{\bf p}^{2}}{90}+\frac{19w\vec{\lambda}^{2}}{10} \biggr) \bigg]\, ,  
\\
-\Gamma _{{\cal A}_{3}^{\prime }}^{(2)}({\bf p}_{1},{\bf p}_{2},\vec{\lambda}
) &=&-w\vec{\lambda}^{2}\frac{u^{1/2}}{\varepsilon }\biggl(\frac{52w\vec{
\lambda}^{2}}{45}+2\frac{{\bf p}_{1}^{2}{\bf +p}_{2}^{2}}{15}\biggr)\, ,
\nonumber \\
\end{eqnarray}
\end{subequations}
where ${\bf p}={\bf p}_{1}{\bf +p}_{2}$, and
\begin{subequations}
\label{resGamma3}
\begin{eqnarray}
-\Gamma _{{\cal A}_{0}^{\prime }}^{(3)}(\{{\bf p}\},\{\vec{\lambda}\}) &=&0\, ,  
 \\
-\Gamma _{{\cal A}_{1}^{\prime }}^{(3)}(\{{\bf p}\},\{\vec{\lambda}\}) &=&-
\frac{w\sum_{i=1}^3\vec{\lambda}_{i}^{2}}{3} \, \frac{u^{3/2}}{2\varepsilon }\, ,  
 \\
-\Gamma _{{\cal A}_{2}^{\prime }}^{(3)}(\{{\bf p}\},\{\vec{\lambda}\}) &=&
\frac{w\sum_{i=1}^3\vec{\lambda}_{i}^{2}}{3}\, \frac{21u^{3/2}}{4\varepsilon }\, ,  
\\
-\Gamma _{{\cal A}_{3}^{\prime }}^{(3)}(\{{\bf p}\},\{\vec{\lambda}\}) &=&
\frac{w\sum_{i=1}^3\vec{\lambda}_{i}^{2}}{3}^{2} \, \biggl(1-\frac{31u}{6\varepsilon }\biggr)\, . \quad
\end{eqnarray}
\end{subequations}
In writing Eqs.~(\ref{resGamma2}) and (\ref{resGamma3}) we have dropped inconsequential factors $(\tau/\mu^2)^{-\varepsilon /2}$ and $(\tau/\mu^2)^{-\varepsilon /2} (\mu^\varepsilon/G_\varepsilon)^{1/2}$, respectively, for notational simplicity.

For the purpose of renormalization it is handy to collect the operators in a vector
\begin{equation}
\underline{{\cal A}^{\prime }}=({\cal A}_{0}^{\prime },\ldots ,{\cal A}_{3}^{\prime })^T \, . 
\end{equation}
Due to the mixing, a proper renormalization requires an entire renormalization matrix $\underline{\underline{Z}}$. We set
\begin{equation}
\label{opRenRRN}
\underline{{\cal A}}^{\prime }\rightarrow \underline{\mathaccent"7017{\cal A}
}^{\prime }\ ,\qquad \underline{{\cal A}}^{\prime }=\underline{\underline{Z}}
\ \underline{\mathaccent"7017{\cal A}}^{\prime } \,  .
\end{equation}
This means in turn that the vertex functions with an insertion of $\underline{{\cal A}^{\prime }}$ are renormalized by
\begin{equation}
\Gamma _{\underline{{\cal A}}^{\prime }}^{(n)}\rightarrow \mathaccent"7017{
\Gamma }_{\underline{\mathaccent"7017{\cal A}}^{\prime }}^{(n)}\ ,\qquad
\Gamma _{\underline{{\cal A}}^{\prime }}^{(n)}=Z^{n/2}\underline{\underline{Z
}} \ \mathaccent"7017{\Gamma }_{\underline{\mathaccent"7017{\cal A}}
^{\prime }}^{(n)}\, .  \label{Matr-Ren}
\end{equation}
The $\varepsilon $-poles are eliminated by minimal subtraction. To this end, we introduce the Laurent expansion of the $\underline{\underline{Z}}$-matrix
\begin{equation}
\underline{\underline{Z}}=\underline{\underline{1}}+\sum_{k=1}^{\infty }
\frac{1}{\varepsilon ^{k}} \, \underline{\underline{M}}^{(k)} .
\end{equation}
Our 1-loop calculation leads to
\begin{equation}
\label{M1}
\underline{\underline{M}}^{(1)}=\left(
\begin{array}{cccc}
0 & 0 & 0 & 0 \\
\frac{u}{90} & -\frac{u}{2} & \frac{u}{15} & \frac{u^{3/2}}{2} \\
-\frac{u}{90} & -\frac{19u}{10} & -\frac{17u}{15} & -\frac{21u^{3/2}}{4} \\
0 & \frac{52u^{1/2}}{45} & \frac{4u^{1/2}}{15} & \frac{17u}{4}
\end{array} \right) \,  .
\end{equation}

Bare, unrenormalized quantities are independent of the external length scale
parameter $\mu $. Thus, we have the identity
\begin{equation}
\mu \frac{\partial }{\partial \mu }\mathaccent"7017{\Gamma }^{(n)}_{\underline{
\mathaccent"7017{\cal A}}^{\prime }} \left( \big\{ {\rm{\bf x}} ,\vec{\lambda} \big\} ; \mathaccent"7017{\tau}, \mathaccent"7017{u}, \mathaccent"7017{w}, \mu \right) = \underline{0}
\end{equation}
which translates via the renormalizations (\ref{reno}) and (\ref{Matr-Ren}) into the
RG equation
\begin{eqnarray}
\label{specificRGE}
&&\biggl(\mu \frac{\partial }{\partial \mu }+ \kappa \tau \frac{\partial }{\partial \tau} +\zeta w\frac{\partial }{\partial w
}+\beta \frac{\partial }{\partial u}-\frac{n}{2}\gamma +\underline{
\underline{\gamma }}\biggr) 
\nonumber \\
&& \qquad \times \,  \Gamma_{\underline{{\cal A}}^{\prime}}^{(n)} \left( \big\{ {\rm{\bf x}} ,\vec{\lambda} \big\} ; \tau, u, w, \mu \right) = \underline{0} \, .
\end{eqnarray}
The matrix $\underline{\underline{\gamma }}$ is given by
\begin{equation}
\label{defGamma}
\underline{\underline{\gamma }}=-\frac{\varepsilon }{2}\, \underline{\underline{
k}}+u\frac{\partial }{\partial u}\, \underline{\underline{M}}^{(1)}+\frac{1}{2} \left( 
\underline{\underline{k}} \, \underline{\underline{M}}^{(1)} - \underline{\underline{M}}^{(1)}  \underline{\underline{k}}  \right)\,  .
\end{equation}
Here, $\underline{\underline{k}}$ is the diagonal matrix with the diagonal elements $k_{0}$, $k_{1}$, $k_{2}$, and $k_{3}$. To obtain a fixed point solution to the RG equation~(\ref{specificRGE}) we recast  $\underline{\underline{\gamma }}^\ast = \underline{\underline{\gamma }} (u^\ast)$ in terms of its eigenvalues and eigenvectors, 
\begin{equation}
\label{spectral}
\underline{\underline{\gamma }}^\ast = \sum_{m=0}^3 | m \rangle \, \eta_m  \langle m | \, ,
\end{equation}
where $\eta_m $ are the eigenvalues and $ \langle m |$ and $| m \rangle$ are the corresponding left and right eigenvectors. An RG equation for an eigenoperator 
\begin{equation}
{\cal O}_m=  \langle m |  \, \underline{{\cal A}}^{\prime } 
\end{equation}
is then readily derived form the so-obtained fixed point version of Eq.~(\ref{specificRGE}) by multiplication with $\langle m |$, 
\begin{eqnarray}
\label{specificRGEeigenOp}
&&\biggl(\mu \frac{\partial }{\partial \mu }+ \kappa^\ast \tau \frac{\partial }{\partial \tau} +\zeta^\ast w\frac{\partial }{\partial w
} -\frac{n}{2}\eta + \eta_m \biggr) 
\nonumber \\
&& \qquad \times \,  \Gamma_{{\cal O}_m}^{(n)} \left( \big\{ {\rm{\bf x}} ,\vec{\lambda} \big\} ; \tau, u^\ast, w, \mu \right) = 0 \, .
\end{eqnarray}
Using the method of characteristics, it is straightforward to solve this RG equation. Augmenting its solution with a dimensional analysis to account for naive dimensions we find the scaling form
\begin{eqnarray}
\label{scalingEigenOps}
&&\Gamma^{(n)}_{{\cal O}_m} \left( \left\{ {\rm{\bf x}} ,\vec{\lambda} \right\} ; \tau, u, w, \mu \right) = 
\ell^{-(d-2+\eta)n/2 + x_m} 
\nonumber \\
&& \times \,
\Gamma^{(n)}_{{\cal O}_m} \left( \left\{ \ell{\rm{\bf x}} , \vec{\lambda} \right\} ; \ell^{-1/\nu}\tau , u^\ast,  \ell^{-\phi /\nu}w, \mu \right) 
\end{eqnarray}
with the scaling dimensions $x_{m}$ of the eigenoperator ${\cal O}_m$ given by
\begin{eqnarray}
x_{m}=d+2+\eta_m\, . 
\end{eqnarray}

At this stage our scaling solution~(\ref{scalingEigenOps}) is rather formal, and we still have to determine the eigenvalues of $\underline{\underline{\gamma }}^\ast$.  From Eqs.~(\ref{M1}) and (\ref{defGamma}) we obtain
\begin{equation}
\underline{\underline{\gamma }}=\left(
\begin{array}{cccc}
0 & 0 & 0 & 0 \\
\frac{u}{90} & -\frac{u}{2} & \frac{u}{15} & \frac{u^{3/2}}{2} \\
-\frac{u}{90} & -\frac{19u}{10} & -\frac{17u}{15} & -\frac{21u^{3/2}}{4} \\
0 & \frac{52u^{1/2}}{45} & \frac{4u^{1/2}}{15} & \frac{17u}{4} -\frac{\varepsilon}{2}
\end{array}
\right)  .
\end{equation}
At the infrared stable fixed point $u^{\ast}$ this matrix has the eigenvalues
\begin{subequations}
\begin{eqnarray}
\label{resEta0}
\eta_0 &=&0\, ,  
 \\
\label{resEta1}
\eta_1 &=&-\zeta_{\ast }=\frac{\phi }{\nu }-2\, ,
\\
\eta_2 &=&\Big(23-8\sqrt{30}\Big)\frac{
\varepsilon }{105}+O(\varepsilon ^{2}) \, ,  
\\
\eta_3 &=&\Big(23+8\sqrt{30}\Big)\frac{
\varepsilon }{105}+O(\varepsilon ^{2}) \, .
\end{eqnarray}
\end{subequations}
The left eigenvector belonging to $\eta_0$ reads $\langle 0 |=(1,0,0,0)$, i.e., ${\cal A}_{0}$ is an eigenoperator of the RG and its scaling behavior is governed by $\eta_0$. From Sec.~\ref{eom} we know rigorously that the scaling dimension of ${\cal A}_{0}$ is $x_0 = d+2$. Hence, Eq.~(\ref{resEta0}) holds to arbitrary order in $\varepsilon$ expansion. $\eta_1$ is associated with the left eigenvector $\langle 1 |=(0,2,2,3(u^{\ast })^{1/2})$. Thus, we can identify its eigenoperator with ${\cal B}_{1} \cdot \mathcal{H}^\prime$, cf.\ Sec.~\ref{eom}, and conclude that Eq.~(\ref{resEta1}) is valid to arbitrary order in $\varepsilon$ expansion. To 1-loop order $\eta_1$ is given by $\eta_1 =  - \frac{4\varepsilon }{21}+O(\varepsilon ^{2})$.

We observe that our 1-loop results for $\eta_0$ and $\eta_1$ are in full agreement with our non-perturbative results deduced from the equation of motion. In other words, our 1-loop calculation fulfills important stringent consistency checks. Of course, these checks do not guarantee the correctness of our results for $\eta_2$ and $\eta_3$. However, they reassure us that important features of our calculations like, e.g., symmetry factors, are correct. In this indirect sense, the consistency checks are in favor of our results for $\eta_2$ and $\eta_3$.

To extract the sought-after corrections to scaling we need to know the scaling behavior of the coupling constants associated with the RG eigenoperators.  The renormalized contribution of the
eigenoperators to the Hamiltonian ${\cal H}$ reads $\int
d^{d}x\,\sum_{m }v_{m}{\cal O}_m$. The flow of the
coupling constants $v_{m}$ under renormalization is therefore described by $\bar{v}_{m }(l)=l^{\omega _{m}}v_{m}$ with the correction
to scaling exponents
\begin{equation}
\omega _{m}=x_{m }-d=2+\eta_m\, .
\end{equation}
After all, we find these exponents stemming from the RRN specific irrelevant
operators to be given by
\begin{subequations}
\label{expres}
\begin{eqnarray}
\omega _{0} &=& 2\, ,
\\
\omega _{1}&=&\phi /\nu \, ,
\\
\omega _{2} &=& 2-0.198\,\varepsilon +O(\varepsilon ^{2})\, ,
\\
\omega_{3} &=& 2+0.636\,\varepsilon +O(\varepsilon ^{2})\, .
\end{eqnarray}
\end{subequations}

Before we turn to the general percolation operators, we find it instructive to briefly re-analyze the work on corrections to scaling in RRN by HL. If one
erroneously neglects the coupling of the operator ${\cal A}_{1}$ to the
other operators under renormalization one is led to a "scalar"
renormalization factor $Z_{HL}$ that corresponds to the matrix element $Z_{1,1}$ of $\underline{\underline{Z}}$. Thus, one would find to 1-loop order
\begin{eqnarray}
{\cal A}_{1} &\rightarrow &\mathaccent"7017{\cal A}_{1}=Z_{HL}^{-1}{\cal A}_{1}\, ,  
\\
Z_{HL} &=&1+\frac{1}{\varepsilon }M_{1,1}^{(1)}+O(u^{2})=1-\frac{u}{2\varepsilon }+O(u^{2})\, ,  \qquad
\\
\gamma _{HL} &=&-\frac{u}{2}+O(u^{2})\, .
\end{eqnarray}
The anomalous dimension were then $\eta _{HL}=\gamma _{HL\ast }=-\varepsilon /7+O(\varepsilon ^{2})$, which results in an exponent $\omega_{HL}=2+\eta _{HL}=2-\varepsilon /7+O(\varepsilon ^{2})$. The corresponding crossover exponent $\phi_2$ defined by HL follows as $\phi_{2}=2\phi -\nu \omega_{HL}=(2-\frac{5\varepsilon }{21})^{-1}[2(2-\frac{4\varepsilon }{21})-(2-
\frac{\varepsilon }{7})]=1+O(\varepsilon ^{2})$. Our analysis above shows clearly that this exponent has no meaning as far as corrections to scaling in RRN are concerned.

\subsection{Spin specific operators}
To analyze corrections to scaling in continuous spin systems, we have to take into account the operator ${\cal A}_{c}$ defined in Eq.~(\ref{Ac}). This operator breaks the $O (D)$ symmetry in replica space. In this sense, its symmetry is lower than that of the resistor specific operators. This lower symmetry has an important consequence. Though the renormalization of ${\cal A}_{c}$ generates  the operators ${\cal A}_{0}^{\prime },\ldots ,{\cal A}_{3}^{\prime}$, these do not in turn generate ${\cal A}_{c}$ under renormalization. 

This structure is very similar to the one we encountered in studying multifractality in RRN~\cite{stenull_janssen_epl_2000,stenull_janssen_2001} and random resistor diode networks~\cite{stenull_janssen_epl_2001,stenull_janssen_pre_2002}. In the field theoretic description of these networks, there are dangerous irrelevant operators corresponding to the multifractal moments of the current distribution on the networks. These operators (masters) generate a whole bunch of other operators (servants). The servants on the other hand, do not generate their masters. All servants must be taken into account  in the renormalization process, at least in principle. However, the renormalization matrices have a particular, simple structure. Due to this simple structure, the scaling index of a master operator is completely determined by a single element of the renormalization matrix. Hence, for the practical purpose of calculating a masters scaling index, the servants can be neglected.

To facilitate the renormalization of the spin specific operators, we introduce the vector
\begin{equation}
\underline{{\cal A}}_{\text{ss}}=({\cal A}_c, {\cal A}_{0}^{\prime },\ldots ,{\cal A}_{3}^{\prime })^T \, . 
\end{equation}
A proper renormalization requires a $5\times 5$ renormalization matrix $\underline{\underline{Z}}_{\text{ss}}$ which we introduce by setting
\begin{equation}
\underline{{\cal A}}_{\text{ss}}\rightarrow \underline{\mathaccent"7017{\cal A}
}_{\text{ss}}\ ,\qquad \underline{{\cal A}}_{\text{ss}}=\underline{\underline{Z}}_{\text{ss}}
\,  \underline{\mathaccent"7017{\cal A}}_{\text{ss}} \,  .
\end{equation}
The arguments given above imply that $\underline{\underline{Z}}_{\text{ss}}$ is of the form
\begin{equation}
\label{Zss}
\underline{\underline{Z}}_{\text{ss}}=
\left(
\begin{array}{cc}
Z_c & \underline{X}_c \\
\underline{0}^T & \underline{\underline{Z}}
\end{array} \right)   ,
\end{equation}
where $ \underline{\underline{Z}}$ is the renormalization matrix defined in Eq.~(\ref{opRenRRN}), and $\underline{0} = (0,0,0,0)$. The elements of $\underline{X}_c$ must be chosen, in principle, to cancel $\varepsilon$ poles associated with the servants generated by ${\cal A}_c$. In practice, however, we do not need to determine these elements for computing the correction to scaling exponents. In a 1-loop calculation we find
\begin{equation}
\label{Zc}
Z_{c}=1+\frac{17u}{30\varepsilon }+O(u^{2}) \, .
\end{equation}
The matrix $\underline{\underline{\gamma}}_{\text{ss}}$, which is given up to an obvious modification by Eq.~(\ref{defGamma}), inherits the simple structure of $\underline{\underline{Z}}_{\text{ss}}$. As a consequence, the eigenvalues of $\underline{\underline{\gamma}}_{\text{ss}}$ at the fixed point $u^\ast$ are $\eta_0$ to $\eta_3$ and $\eta_c. $ From Eq.~(\ref{Zc}) we obtain  
\begin{equation}
\eta_c =- \frac{17\varepsilon}{105}+O(\varepsilon^{2}) \, .
\end{equation}
This eigenvalue leads finally to a correction exponent
\begin{equation}
\omega_{c}=2-\frac{17\varepsilon }{105}+O(\varepsilon ^{2})\, .
\end{equation}
in addition to $\omega_0, \cdots, \omega_3$. The crossover exponent $\phi _{c}$ of HL is related to  $\omega_c$ via $\phi _{c}=2\phi -\nu \omega _{c}$. We obtain to 1-loop order 
\begin{equation}
\phi_{c}=1+\frac{\varepsilon }{105}+O(\varepsilon ^{2})
\end{equation}
 in conformity with the value given by HL.

\section{Renormalization of general percolation operators}
\label{gpo}
Now we turn to the renormalization of the operators $\mathcal{A}_4$ to $\mathcal{A}_7$. The corrections to scaling arising from these operators have been calculated a long time ago by Amit {\em et al}.~\cite{amit_wallace_zia_77}. Nevertheless, we think that the general percolation operators deserve some attention here for 2 reasons: (i) Amit {\em et al}.\ used the usual Potts model formulation of percolation that is different from the RRN formulation in the way constraints on the order parameter field are implemented, viz.\ $\sum_{\vec{\theta}} s (\brm{x}, \vec{\theta}) = 0$ versus  $ \psi_{\vec{0}} (\brm{x})= 0$. It seems desirable to have a treatment of both the specific and the general percolation operators that is self-contained within one formulation. (ii) Because the calculation is somewhat involved, two independent approaches help to guarantee the correctness of results. In fact, our results concerning the general percolation operators coincide in the end with the results of Ref.~\cite{amit_wallace_zia_77}.

\begin {figure}[ptb]
\includegraphics [width=7.4cm]{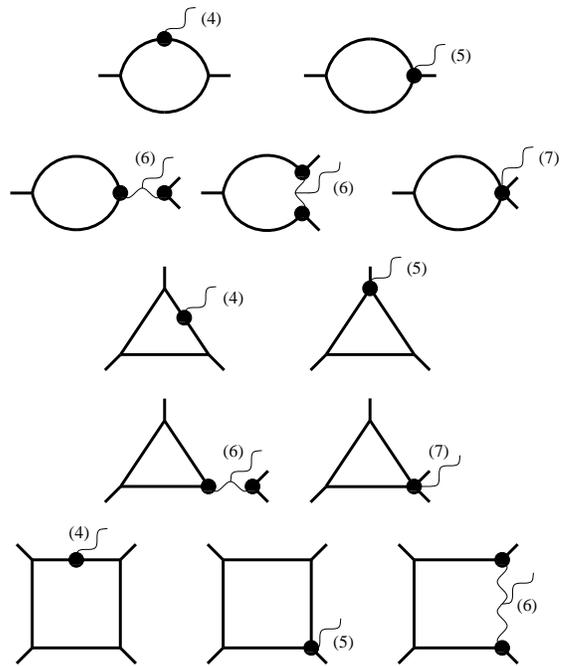}\caption {Diagrams contributing at 1-loop order to the renormalization of the general percolation operators. The bold lines symbolize bold propagators. Single dots accompanied by a wiggly line respectively stand for an insertion of (4) ${\cal A}_4^\prime$, (5) ${\cal A}_5^\prime$, and (7) ${\cal A}_7^\prime$. An insertion of ${\cal A}_6^\prime$ is depicted with help of 2 dots and 2 wiggly lines. Note that we have, in contrast to Fig.~\ref{fig1} displayed bold instead of conducting diagrams to save space.}
\label{fig2}
\end {figure}

The general percolation operators affect the renormalization of the 2, 3 and 4-point vertex functions. To asses this effect, we have to calculate these vertex functions with insertions of $\mathcal{A}_4$ to $\mathcal{A}_7$. The Feynman diagrams that contribute to these vertex functions are summarized in Fig.~\ref{fig2}. To calculate these diagrams, we decompose them into their conducting diagrams and then take the limit $w \to 0$. Then, the conducting and the insulating propagators can be replaced by simple $1/(\tau + \brm{p}^2)$ propagators. For most of the diagrams, decomposition culminates into to a simple numerical factor. A little caution must be exercised, however, to distinguish between terms corresponding to $\mathcal{A}_6$ and $\mathcal{A}_7$, respectively. To foster this distinction, we keep the dependence of these two operators on the replica currents explicit. This dependence is different in the way the sum over the external replica currents, say $\vec{\lambda}_1, \cdots , \vec{\lambda}_4$ vanishes, viz.\ $\mathcal{A}_6$ is proportional to $S = (\delta_{\vec{\lambda}_1+\vec{\lambda}_2 , \vec{0}} \, \delta_{\vec{\lambda}_3+\vec{\lambda}_4 ,\vec{0}} + \delta_{\vec{\lambda}_1+\vec{\lambda}_3 ,\vec{0}} \, \delta_{\vec{\lambda}_2+\vec{\lambda}_4 ,\vec{0}} + \delta_{\vec{\lambda}_1+\vec{\lambda}_4 , \vec{0}} \, \delta_{\vec{\lambda}_2+\vec{\lambda}_3 , \vec{0}})/3$ whereas $\mathcal{A}_7$ is proportional to $F = \delta_{\vec{\lambda}_1+\vec{\lambda}_2 +\vec{\lambda}_3+\vec{\lambda}_4, \vec{0}}$.

As for the specific operators, it is convenient to rescale the general operators. To be specific, we rescale the general percolation operators according to Eq.~(\ref{rescaledOps}) with $k_4 =0$, $k_5 =1$ and $k_6 = k_7 = 2$. 

Our 1-loop calculation gives for the 2-point functions with insertions
\begin{subequations}
\label{resGamma2GP}
\begin{eqnarray}
-\Gamma _{{\cal A}_{4}^{\prime }}^{(2)}(\{{\bf p}\}) &=& \brm{p}^4 \, ,
\\
-\Gamma _{{\cal A}_{5}^{\prime }}^{(2)}(\{{\bf p}\}) &=& - \brm{p}^4 \, \frac{u^{1/2}}{9 \, \varepsilon}\, ,
\\
-\Gamma _{{\cal A}_{6}^{\prime }}^{(2)}(\{{\bf p}\}) &=& 0 \, ,
 \\
-\Gamma _{{\cal A}_{7}^{\prime }}^{(2)}(\{{\bf p}\}) &=& 0 \, ,
\end{eqnarray}
\end{subequations}
where we have dropped inconsequential factors $(\tau/\mu^2)^{-\varepsilon /2}$ for notational simplicity. For the 3-point functions with insertions we find
\begin{subequations}
\label{resGamma3GP}
\begin{eqnarray}
-\Gamma _{{\cal A}_{4}^{\prime }}^{(3)}(\{{\bf p}\}) &=& - \frac{\sum_{i=1}^3 \brm{p}_i^2 }{3} \, \frac{2 \, u^{3/2}}{\varepsilon} \, ,  
 \\
-\Gamma _{{\cal A}_{5}^{\prime }}^{(3)}(\{{\bf p}\}) &=& \frac{\sum_{i=1}^3 \brm{p}_i^2 }{3}  \left( 1 - \frac{2 \, u}{3 \, \varepsilon} \right) \, ,  
 \\
-\Gamma _{{\cal A}_{6}^{\prime }}^{(3)}(\{{\bf p}\}) &=& \frac{\sum_{i=1}^3 \brm{p}_i^2 }{3} \,  \frac{u^{1/2}}{3\, \varepsilon} \, ,  
\\
-\Gamma _{{\cal A}_{7}^{\prime }}^{(3)}(\{{\bf p}\}) &=& - \frac{\sum_{i=1}^3 \brm{p}_i^2 }{3}  \, \frac{u^{1/2}}{2\, \varepsilon}\, .
\end{eqnarray}
\end{subequations}
Here, we omitted factors $(\tau/\mu^2)^{-\varepsilon /2} (\mu^\varepsilon/G_\varepsilon)^{1/2}$. Our results for the 4-point functions with insertions read
\begin{subequations}
\label{resGamma4GP}
\begin{eqnarray}
-\Gamma _{{\cal A}_{4}^{\prime }}^{(4)}(\{{\bf p}\}) &=& S \, \frac{24\, u^2}{\varepsilon} - F\, \frac{36\, u^2}{\varepsilon} \, ,  
 \\
-\Gamma _{{\cal A}_{5}^{\prime }}^{(4)}(\{{\bf p}\}) &=& - S \, \frac{16\, u^{3/2}}{\varepsilon} + F\, \frac{24\, u^{3/2}}{\varepsilon} \, ,  
 \\
-\Gamma _{{\cal A}_{6}^{\prime }}^{(4)}(\{{\bf p}\}) &=& S \left( 1 - \frac{6\, u}{\varepsilon} \right) + F\, \frac{4\, u}{\varepsilon} \, ,  
\\
-\Gamma _{{\cal A}_{7}^{\prime }}^{(4)}(\{{\bf p}\}) &=&  S \, \frac{6\, u}{\varepsilon} + F \left( 1 - \frac{12\, u}{\varepsilon} \right) \, ,
\end{eqnarray}
\end{subequations}
where we dropped factors $(\tau/\mu^2)^{-\varepsilon /2} \mu^\varepsilon/G_\varepsilon$. 

The remaining RG analysis proceeds completely analogous to that for the RRN specific operators. The RG equation governing the general percolation operators has the same structure as Eq.~(\ref{specificRGE}). Of course, here we can set $w=0$. Basically, we just need to make the replacements
\begin{eqnarray}
\underline{\mathcal{A}}^\prime &\to& \underline{\mathcal{A}}_{\text{gp}}^{\prime}  = (\mathcal{A}_{4}^{\prime },\ldots ,\mathcal{A}_{7}^{\prime }) \, ,
\\
\underline{\underline{\gamma }} &\to& \underline{\underline{\gamma}}_{\text{gp}} \, ,
\end{eqnarray}
with the latter matrix resulting from (\ref{resGamma2GP}), (\ref{resGamma3GP}) and (\ref{resGamma4GP}) as
\begin{equation}
\underline{\underline{\gamma }}_{\text{gp}} =\left(
\begin{array}{cccc}
-\frac{u}{6} & 2 u^{3/2} & -24 u^2 & 36 u^2 \\
\frac{u^{1/2}}{9} & \frac{5u}{12} -\frac{\varepsilon}{2} & 16 u^{3/2} & -24 u^{3/2} \\
0 & -\frac{u^{1/2}}{3} & \frac{17 u}{3} - \varepsilon& -4 u \\
0 & \frac{u^{1/2}}{2} & -6 u &  \frac{35 u}{3} - \varepsilon
\end{array}
\right) \, .
\end{equation}
At the infrared stable fixed point, $\underline{\underline{\gamma }}_{\text{gp}}$ has the eigenvalues
\begin{subequations}
\label{eigenvaluesGP}
\begin{eqnarray}
\label{eigenvalue4}
\eta_4 &=& 0 \, ,  
 \\
\label{eigenvalue5} 
\eta_5 &=& - \eta\, ,  
 \\
\eta_6 &=& \left( 26 - \sqrt{889} \right)\frac{\varepsilon}{21} + O (\varepsilon^2) \, ,  
\\
\eta_7 &=&\left( 26 + \sqrt{889} \right)\frac{\varepsilon}{21} + O (\varepsilon^2) \, .
\end{eqnarray}
\end{subequations}
The eigenvalues $\eta_4$ and $\eta_5$ are associated with the left eigenvectors $\langle 4 | = (2, 3(u^\ast)^{1/2}, 0, 0)$ and $\langle 5 | = (1, 3(u^\ast)^{1/2}, -3u^\ast, 3u^\ast)$, respectively. By comparison with the $w\to 0$ limit of Eqs.~(\ref{Eq.Mot.}), we identify ${\cal B}_{2} \cdot \mathcal{H}^\prime$ and ${\cal B}_{3} \cdot \mathcal{H}^\prime$ as the corresponding eigenoperators. It follows that Eqs.~(\ref{eigenvalue4}) and (\ref{eigenvalue5}) are correct to arbitrary order in perturbation theory. The eigenvalues stated in Eqs.~(\ref{eigenvaluesGP}) entail the correction to scaling exponents
\begin{subequations}
\label{correctionGP}
\begin{eqnarray}
\omega_4 &=& 2 \, ,  
 \\
\omega_5 &=& 2 - \eta \, ,  
 \\
\omega_6 &=& 2 - 0.182 \varepsilon+ O (\varepsilon^2)\, ,  
\\
\omega_7 &=&  2 + 2.658 \varepsilon + O (\varepsilon^2) 
\end{eqnarray}
\end{subequations}
in full agreement with the results of Amit {\em et al}.

\section{Critical behavior of the average resistance etc.}
\label{scalingResults}
Now we can harvest the results of our RG analysis to write down scaling expressions, including the most important corrections, for key observables for the RRN and the $x$-$y$-model. We will elaborate on the RRN in some detail and assemble step by step the generating function $G(\brm{x}, \brm{x}^\prime, \vec{\lambda})$ from which we then extract $M_R^{(1)}$ and $\Sigma$. To cover spin models, the generating function requires some modifications that will be explained.

\subsection{Random resistor networks}
The generating function has two kinds of ingredients, viz.\ the 2-point correlation function without insertions and the set of 2-point correlation functions obtained by inserting each of the RG eigenoperators. From Eq.~(\ref{scaling}), Eq.~(\ref{scalingEigenOps}) as well as its counterpart for the general percolation operators we collect that 
\begin{widetext}
\begin{eqnarray}
\label{sammlung}
G\left(\brm{x}, \brm{x}^\prime, \vec{\lambda}\right) = \ell^{d-2+\eta} \bigg\{ G^{(2)} \left( \ell |\brm{x} - \brm{x}^\prime|, \ell^{-\phi/\nu} w \vec{\lambda}^2\right) 
+  \sum_{k=1}^7 v_k \, \ell^{\omega_k} \, G^{(2)}_{\mathcal{O}_k} \left( \ell |\brm{x} -\brm{x}^\prime|, \ell^{-\phi/\nu} w \vec{\lambda}^2\right) + \cdots \bigg\} . \qquad
\end{eqnarray}
Here, we used that the 2-point correlation function is the inverse of the 2-point vertex function and we applied Fourier transformation to switch from momentum to position space. The ellipsis stands for contribution from irrelevant operators with a naive dimension higher than 8. Next we expand Eq.~(\ref{sammlung}) in a power series in $w \vec{\lambda}^2$ as well as in the deviation $u -u^\ast$ from the fixed point. Choosing $\ell = |\brm{x} - \brm{x}^\prime|^{-1}$ and setting all non-universal constants equal to 1 for notational simplicity we get the leading terms
\begin{eqnarray}
\label{sammlungExpanded}
&& G\left(\brm{x}, \brm{x}^\prime, \vec{\lambda}\right) =  |\brm{x} - \brm{x}^\prime|^{-(d-2+\eta)} \bigg\{
  \Big[ 1 + w \vec{\lambda}^2  |\brm{x} - \brm{x}^\prime|^{\phi/\nu} + \cdots \Big]
 \Big[ 1 + [u -u^\ast]  |\brm{x} - \brm{x}^\prime|^{-\omega} + \cdots \Big] \qquad
\nonumber \\
&& + \, \sum_{k=1}^3 w \vec{\lambda}^2  |\brm{x} - \brm{x}^\prime|^{-\omega_k + \phi/\nu} + \cdots
 + \sum_{k=4}^7 |\brm{x} - \brm{x}^\prime|^{-\omega_k} \Big[ 1 + w \vec{\lambda}^2  |\brm{x} - \brm{x}^\prime|^{\phi/\nu} + \cdots \Big] + \cdots \bigg\} \, . 
\end{eqnarray}
\end{widetext}
By virtue of Eq.~(\ref{corrPsiRRN}) we know that we now can extract the average resistance by taking the derivative with respect to $-\vec{\lambda}^2/2$ and then setting $\vec{\lambda} = \vec{0}$. After all, we arrive at
\begin{eqnarray}
\label{sumRes}
M_{R}^{(1)} &\sim&  |x-x^{\prime }|^{\phi /\nu }\, \bigg[ 1+A \, |x-x^{\prime }|^{-\omega } 
\nonumber \\
&+& \sum_{k=1}^7 A_k \, |x-x^{\prime }|^{-\omega_k } + \cdots \bigg] \, ,
\end{eqnarray}
Here, $A$ and $A_k$ are non-universal constants. We opted to write them down explicitly here to make closer contact to Eq.~(\ref{wrong}). $A$ goes to zero for $u\to u^\ast$. The $A_k$ vanish for $v_k \to 0$.

Next we turn to the average conductivity of the RRN. Commonly, the conductivity $\Sigma$ of percolating systems is defined with respect to a bus bar geometry where the network is placed between two parallel superconducting plates (the electrodes) of area $L^{d-1}$ a distance $L$ apart. From the above we expect that the average conductance $\sigma$ of this finite-size system scales as
\begin{eqnarray}
\label{scaleFormConductance2}
\sigma (L, \tau) &=& \left| \tau \right|^{\phi} \bigg[ \Pi \left( L/\xi \right) + \Xi \left( L/\xi \right) \, |\tau|^{\nu \omega} 
\nonumber \\
&+& \sum_{k=1}^7 \Xi_k  \left( L/\xi \right) \, |\tau|^{\nu \omega_k}  + \cdots \bigg] \, ,
\end{eqnarray}
where  $\xi \sim |\tau|^{-\nu}$ is the percolation correlation length and  $\Pi$, $\Xi$ and the $\Xi_k$ are scaling function with the properties
\begin{eqnarray}
\label{propsPi}
\Pi (x) \sim \Xi (x) \sim \Xi_k (x) \sim
\left\{ 
\begin{array}{ccc}
const &\mbox{for}& x \ll 1 \\
x^{d-2} &\mbox{for}& x \gg 1
\end{array}
\right. \, .
\end{eqnarray}
Above the percolation threshold ($\tau <0$) the RRN behaves on length scales large compared to the correlation length $\xi \sim |\tau|^{-\nu}$ like a homogeneous system with conductivity $\Sigma$. Hence, we may write for $L\gg \xi$ that
\begin{eqnarray}
\label{zwiSigma}
\Sigma (\tau) \sim L^{2-d} \sigma (L, \tau) \, .
\end{eqnarray}
Merging Eqs.~(\ref{scaleFormConductance2}), (\ref{propsPi}) and (\ref{zwiSigma}) we finally get
\begin{eqnarray}
\label{sumResCond}
\Sigma \sim |\tau|^{t}\, \bigg[ 1+B  \, |\tau|^{\nu \omega } 
+ \sum_{k=1}^7 B_k \, |\tau|^{\nu \omega_k } + \cdots \bigg] \, ,
\end{eqnarray}
with non-universal amplitudes $B$ and $B_k$. Of course, $B$ vanishes for $u\to u^\ast$ and the $b_k$ vanish for $v_k \to 0$.

\subsection{Spin systems}
For spin systems like the $x$-$y$-model we had to include the irrelevant operator $\mathcal{A}_c$ into our RG analysis because these systems are, unlike the RRN, not $O(D)$-invariant in replica space. Thus, the RG has 9 irrelevant eigenoperators of naive dimension 8 and the counterpart of Eq.~(\ref{sammlung}) has an extra term that features the exponent $\omega_c$. By applying basically the same steps as for the average resistance we find that the first cumulant of the angular fluctuations scales as  
\begin{eqnarray}
\label{sumResC}
C_{\varphi}^{(1)} &\sim&  |x-x^{\prime }|^{\phi /\nu }\, \bigg[ 1+C \, |x-x^{\prime }|^{-\omega } 
 \\
&+& C_c \, |x-x^{\prime }|^{-\omega_c} + \sum_{k=1}^7 C_k \, |x-x^{\prime }|^{-\omega_k } + \cdots \bigg] \, ,
\nonumber
\end{eqnarray}
where $C$, $C_c$ and the $C_k$ are non-universal amplitudes.

Equation~(\ref{sumResC}) concludes our results on corrections to scaling. However, our analysis also sheds light on the  second cumulant of the angular fluctuations. Equation~(\ref{corrPsiXY}) implies that we can derive $C_{\varphi}^{(2)}$ via taking the derivative with respect to $K_2 (\vec{\lambda})$. This homogenous polynomial in $\vec{\lambda}$ exclusively appears in $\mathcal{A}_c$. We saw that $\mathcal{A}_c$ has the property of being a master operator whose scaling behavior is governed by $\omega_c$. Consequentially, we obtain that
\begin{eqnarray}
\label{ResC2}
C_{\varphi}^{(2)} \sim  |x-x^{\prime }|^{\phi_c/\nu} + \cdots \, .
\end{eqnarray}
Here, $\mathcal{A}_c$ gives rise to the leading scaling behavior, i.e.,  $\mathcal{A}_c$ is a dangerous irrelevant operator as far as $C_{\varphi}^{(2)}$ is concerned. 

Of course it is interesting in this context to ask, what the leading scaling behavior of the higher cumulants might be. It can be shown to arbitrary order in perturbation theory~\cite{stenull_janssen_xyMulti}, that
\begin{eqnarray}
\label{ResCgeneral}
C_{\varphi}^{(l)} \sim  |x-x^{\prime }|^{\psi_l/\nu} + \cdots \, ,
\end{eqnarray}
where the $\psi_l$ are the critical exponents of the multifractal moments
\begin{eqnarray}
M_I^{(l)} = \left\langle \sum_b \left( \frac{I_b}{I} \right)^{2l} \right\rangle_C^\prime \sim  |x-x^{\prime }|^{\psi_l/\nu} + \cdots 
\end{eqnarray}
of the current distribution on RRNs, with $I_b$ being the current flowing through bond $b$ and $I$ being the total external current. Note that the HL result $\phi_c = 1 + \varepsilon/105 + O(\varepsilon^2)$ and the $\varepsilon$ expansion result~\cite{park_harris_lubensky_87,stenull_janssen_epl_2000,stenull_janssen_2001} for $\psi_2$ (the entire family of the $\psi_l$ is known to two-loop order~\cite{stenull_janssen_epl_2000,stenull_janssen_2001}) are in full agreement. The upshot here is that Eq.~(\ref{ResC2}) represents merely an instance of the general result~(\ref{ResCgeneral}). 

\section{Concluding remarks}
\label{concludingRemarks}

We have studied corrections to scaling in RRN and continuous spin models. As far as the leading scaling behavior and its leading corrections are concerned, the HL model provides a unified description of both systems. Being interested in next to leading corrections, however, we had to consider distinct sets of irrelevant operators for the two systems.

In both systems, we found the typical mixing of irrelevant operators under renormalization. Thus, we had to compute an entire renormalization matrices. This is the reason why we restricted ourself to considering irrelevant operators with a naive dimension 8. At least in principle, one could analyze higher corrections to scaling originating from irrelevant operators of naive dimension 10, 12 and so on. However, in these cases the renormalization matrices become prohibitively big and their computation and diagonalization requires enormous effort.

One of the spin specific operators, namely ${\cal A}_{c}$, is qualitatively different from the remaining irrelevant operators under consideration. ${\cal A}_{c}$ has the properties of a master operator. Hence, it is sufficient to calculate a single renormalization constant to determine its scaling dimension. This can be done in an elegant way and with moderate effort up to two-loop order~\cite{stenull_janssen_xyMulti}. One has to bear in mind, though, that the corrections to scaling are not only resulting from ${\cal A}_{c}$ but also from the other spin specific as well as from the general percolation operators. Over all, it would require a lot of work to extend our results on corrections to scaling to 2-loop order or to include irrelevant operators of higher naive dimension.

It is interesting to compare the corrections to scaling stemming from the specific and the general percolation operators with the corrections arising from the presence of a surface.  In Ref.~\cite{stenull_janssen_oerding_2001} we studied a semi-infinite RRN at the so-called special and ordinary transitions~\cite{footnote_surface}. We calculated the corrections to $M_R^{(1)}$ induced by the surface when the terminal points $\brm{x}$ and $\brm{x}^\prime$ are located on the surface. At the special transition we found a correction that vanishes as $|\brm{x}-\brm{x}^\prime|^{-\omega_{\mathcal{S}}}$ for increasing terminal separation with the correction to scaling exponent  $\omega_{\mathcal{S}}= 1-\varepsilon/21 + O (\varepsilon^2)$. This correction falls of slower than the corrections induced by the specific and the general percolation operators, i.e., this correction represents the next to leading correction if a surface is present. At the ordinary transition we found a correction $|\brm{x}-\brm{x}^\prime|^{-\omega_{\mathcal{S}}^\infty}$ with $\omega_{\mathcal{S}}^\infty = 3-23\varepsilon/105 + O (\varepsilon^2)$, i.e., this correction vanishes faster than any of the corrections induced by the specific and the general percolation operators.
 
\begin{acknowledgments}
This work has been supported by the Deutsche Forschungsgemeinschaft via the Sonderforschungsbe\-reich 237 ``Unordnung und gro{\ss}e Fluktuationen'' and the Emmy Noether-Programm. 
\end{acknowledgments}

\appendix

\section{Useful identities}
\label{app:ward}
Equation~(\ref{anIdentity}) can be regarded as the centerpiece of Sec.~\ref{eom}. In this appendix we first derive a Ward identity that relates the vertex functions with a single insertion of the operator $\mathcal{H}^\prime$ to the vertex functions without insertion. Then we obtain Eq.~(\ref{anIdentity}) as a corollary by slightly modifying our previous arguments.

\subsection{A Ward identity}
\label{app:ward1}

Now we augment the HL model with an external field, i.e., we consider the Hamiltonian
\begin{eqnarray}
\mathcal{H}_h = \mathcal{H} - (h,s) \, ,
\end{eqnarray}
with $\mathcal{H}$ given by Eq.~(\ref{Hamilt}) with $f_i =0$ and where we have used the abbreviated notation
\begin{eqnarray}
 (h,s) = \int d^d x \int_{\vec{\theta}} \, h (\brm{x},\vec{\theta}) \, s  (\brm{x},\vec{\theta}) \, .
\end{eqnarray}
To facilitate our argument, we consider the shift 
\begin{eqnarray}
s  (\brm{x},\vec{\theta}) \to  s  (\brm{x},\vec{\theta}) + c  (\brm{x},\vec{\theta})
\end{eqnarray}
of the order parameter field. This shift leads to the modification
\begin{eqnarray}
\label{shiftHamil}
\mathcal{H}_h \to \mathcal{H}_h + \left(c, \mathcal{H}_h^\prime \right) 
\end{eqnarray}
of the Hamiltonian. Here and in the following we omit inconsequential terms of order $O ( c^2 )$. By virtue of Eq.~(\ref{shiftHamil}) an application of the shift to
\begin{eqnarray}
\left\langle (h,s)^n \right\rangle = \int \mathcal{D} s \, (h,s)^n \exp \left( - \mathcal{H}_h \right)
\end{eqnarray}
 yields
\begin{equation}
\label{sem}
\left\langle(h,s)^{n}(c,\mathcal{H}_h^{\prime})\right\rangle=n \, (h,c) \,  \left\langle(h,s)^{n-1} \right\rangle\, .
\end{equation}
Dividing both sides of Eq.~(\ref{sem}) by $n!$ and summing from 1 to $\infty$ leads to
\begin{equation}
\label{semmel}
\mathcal{Z}_{(c,\mathcal{H}_h^{\prime})} [h] = (h,c) \, \mathcal{Z} [h] \, ,
\end{equation}
where
\begin{equation}
\mathcal{Z}[h]=\langle\exp(h,s)\rangle=\sum_{n=0}^{\infty}\frac{\langle(h,s)^{n}\rangle}{n!}
\end{equation}
is the generating functional for the correlation functions of the order parameter field and
\begin{equation}
\mathcal{Z}_{(c,\mathcal{H}_h^{\prime})}[h]=\sum_{n=1}^{\infty}\frac{ \langle (c,\mathcal{H}_h^{\prime}) \, \, (h,s)^{n}\rangle}{n!}
\end{equation}
is the generating functional for the corresponding correlation functions with an insertion of $(c,\mathcal{H}_h^{\prime})$. Defining
\begin{equation}
\mathcal{Z}[h,c]= \mathcal{Z}_{(c,\mathcal{H}_h^{\prime})}[h] + \mathcal{Z}_[h]
\end{equation}
and switching to the generating functional of connected correlation functions,
\begin{equation}
\mathcal{W}[h,c]=\ln \left( \mathcal{Z} [h,c] \right) 
\end{equation}
and so on, we find
\begin{equation}
\mathcal{W}[h,c]= \mathcal{W}[h] + (h,c) \, .
\end{equation}
Moving over to the generating functional $\Gamma [h,c]$ of vertex functions via the Legendre transformation
\begin{equation}
\Gamma [s,c] + \mathcal{W}[h,c]= (h,s) \, ,
\end{equation}
with
\begin{equation}
\frac{\delta\Gamma}{\delta s} =h\,,\quad\frac{\delta\mathcal{W}}{\delta
h}=s\,,\quad\frac{\delta\Gamma}{\delta c}=-\frac{\delta\mathcal{W}}{\delta c} \, ,
\end{equation}
we arrive at
\begin{equation}
\Gamma [s,c] = (h,s) -  \mathcal{W}[h] - (h,c) \, .
\end{equation}
Now we take the functional derivative with respect to $c$ to obtain the identity
\begin{equation}
\Gamma_{\mathcal{H}^{\prime}}=\frac{\delta\Gamma}{\delta c}=-\frac
{\delta\Gamma}{\delta s}
\end{equation}
between the generating functional $\Gamma_{\mathcal{H}^{\prime}}$ of vertex functions with an insertion of $\mathcal{H}^{\prime}$ and the generating functional $\Gamma$ of the usual vertex functions with out any insertion. Finally, we take $n$ functional derivatives with respect to the order parameter field to obtain the Ward identity
\begin{equation}
\label{wardHPrime}
\Gamma^{(n)}_{\mathcal{H}^{\prime}}=-\Gamma^{(n+1)}\, .
\end{equation}

\subsection{Derivation of Eq.~(\ref{anIdentity})}
In order to derive Eq.~(\ref{anIdentity}), we just need to modify our arguments of Appendix~(\ref{app:ward1}) slightly. Here, we assume that $c$ is a composite field comprising the order parameter field and possibly its gradients in real and in replica space. Consequentially, we have to replace Eq.~(\ref{sem}) by
\begin{equation}
\left\langle(h,s)^{n}(c,\mathcal{H}_h^{\prime})\right\rangle=n \,  \left\langle (h,c) \, (h,s)^{n-1} \right\rangle\, ,
\end{equation}
where we once more have omitted terms of order $O (c^2)$. Upon taking $n$ functional derivatives with respect to the external field we obtain
\begin{eqnarray}
& & \left\langle s({\bf x}_{1})\cdots s({\bf x}
_{n}) \, (c,\mathcal{H}^{\prime})\right\rangle = 
\\
& & = \,  \sum_{i=1}^{n} \left\langle s({\bf x}_{1})\cdots s({\bf x}_{i-1}) \, c ({\bf x}_{i}) \, s({\bf x}_{i+1})\cdots s({\bf x}_{n})\right\rangle \,  .
\nonumber 
\end{eqnarray}
Setting $c  (\brm{x},\vec{\theta}) = \alpha (\brm{x}, \vec{\theta}) \mathcal{F}  (\brm{x},\vec{\theta})$, where $\alpha$ is independent of the order parameter field, and by taking the functional derivative with respect to $\alpha$ we obtain
\begin{eqnarray}
& & \left\langle s({\bf x}_{1},\vec{\theta}_1)\cdots s({\bf x}
_{n},\vec{\theta}_n) \, \mathcal{F} (\brm{x},\vec{\theta})  \mathcal{H}^{\prime} (\brm{x},\vec{\theta}) \right\rangle 
\nonumber \\
& &= \, \sum_{i=1}^{n}\delta ({\bf x}-{\bf x}_{i}) \delta \big( \vec{\theta} - \vec{\theta}_i \big) \Big\langle s({\bf x}_{1},\vec{\theta}_1)\cdots s({\bf x}_{i-1},\vec{\theta}_{i-1}) 
\nonumber \\
&&\times \,  \mathcal{F}({\bf x}_{i},\vec{\theta}_i)  \, s({\bf x}_{i+1},\vec{\theta}_{i+1})\cdots s({\bf x}_{n},\vec{\theta}_{n})\Big\rangle \,  . \quad
\end{eqnarray}
Equation~(\ref{anIdentity}) is now readily obtained by integrating over $\vec{\theta}$.

\section{Composite fields in the HL- and the Potts model}
\label{app:composite}
In Sec.~\ref{eom} we exploited information on lower dimensional operators to draw conclusions on the scaling dimensions of several eight-dimensional operators that are associated with corrections to scaling. This appendix is intended to provide some background on the RG behavior of the lower dimensional operators and to establish some of the results that serve as an input in Sec.~\ref{eom}.

The following arguments refer primarily to the HL model. Since the HL model reduces to the usual Potts model with $N = (2M)^D$ states upon setting $w=0$, however, our reasonings also apply to the latter model. Being interested in the Potts model, one basically just has to set $w=0$ in any of the formulas in this section. To make closer contact to the conventional notation for the Potts model, one may replace $s (\brm{x},\vec{\theta})$ by $s_i (\brm{x})$ with $i = 1, \cdots, N$ (along with $h (\brm{x},\vec{\theta}) \to h_i (\brm{x})$ for the external field and so on) and the integral $\int_{\vec{\theta}}$ by a summation over $i$.

\subsection{Four-dimensional operators}
We start by considering composite operators with the naive dimension four. The simplest operators of this kind are $\mathcal{B}_0$, $\mathcal{B}_1$ and $\mathcal{B}_2$. The scaling dimensions of these operators follow directly from known RG results. We will revisit these operators briefly towards the end of this subsection. 

$\mathcal{B}_0$ belongs to the trivial representation of the permutation symmetry group. Its counterpart belonging to the fundamental representation is
\begin{eqnarray}
\label{opDef}
\mathcal{C}_1 = s^{2}- \frac{1}{N}\int_{\vec{\theta}}s^{2}\, . 
\end{eqnarray}
The scaling dimension of this operator does not follow immediately from known results. It will be derived in the following. We consider the Hamiltonian
\begin{eqnarray}
\mathcal{H}_\sigma &=&\int d^{d}x\,\sum_{\vec{\theta}}\biggl\{ \frac{\tau }{2}
s^{2}+\frac{1}{2} \sigma s^2 + \frac{1}{2}(\nabla s)^{2}+\frac{w}{2}(\nabla _{\theta }s)^{2}
\nonumber \\
&& +\, \frac{g}{ 6}s^{3} - h s \biggr\}\, ,  \label{HamiltSigma}
\end{eqnarray}
where $h = h (\brm{x},\vec{\theta})$ is an external field and $\sigma = \sigma (\brm{x},\vec{\theta})$ has the property $\int_{\vec{\theta}} \sigma =0$. Due to the extra terms, we need renormalizations in addition to those specified in the renormalization scheme~(\ref{reno}), viz.,
\begin{subequations}
\begin{align}
\label{sigmaRen}
\sigma & \rightarrow \mathring{\sigma}=Z^{-1}Z_{\sigma}\sigma \, ,
\\
h &  \rightarrow \mathring{h}=Z^{-1/2}\Big(h +A\sigma^{2}+B\tau\sigma+C\nabla^{2}\sigma +D w \nabla^{2}_\theta \sigma \Big)\,.
\end{align}
\end{subequations}
In order to determine the renormalization factor $Z_\sigma$ and the additive renormalization constants $A$ to $D$, we perform a shift
\begin{eqnarray}
\label{shiftNeu}
s = s^\prime + c \, ,
\end{eqnarray}
where $c$ is assumed to satisfy the condition $\int_{\vec{\theta}} c =0$. This shift transforms the Hamiltonian so that
\begin{align}
\mathcal{H}_\sigma [s,\tau,\sigma,w,g,h]&=\mathcal{H}_\sigma [s^{\prime},\tau,\sigma^{\prime
},w,g,h^{\prime}]
\nonumber \\
&+\mathcal{H}_\sigma [c,\tau,\sigma,w,g,h]\,,
\end{align}
where
\begin{subequations}
\begin{align}
\label{sigmaPrime}
\sigma^{\prime} & =\sigma+gc\,,
\\
\label{hPrime}
h^{\prime}  &  =h-\tau c-\sigma c+\nabla^{2}c +w \nabla^{2}_\theta c - \frac{g}{2}c^{2} \, .
\end{align}
\end{subequations}
In other words: the Hamiltonian $\mathcal{H}_\sigma$ is form invariant under the shift~(\ref{shiftNeu}) up to an inconsequential term that does not depend on the order parameter field. Note that the shift neither modifies $\tau$ nor $w$ or $g$. Since the $Z$ factors depend only on the dimensionless variant $u$ of $g$, it follows that none of the $Z$ factors is affected by the shift.

Equation~(\ref{shiftNeu}) implies that the renormalized version of $c$ is given by
\begin{eqnarray}
c = Z^{-1/2} \mathring{c} \, .
\end{eqnarray}
Exploiting this we find by renormalizing Eq.~(\ref{sigmaPrime}) that
\begin{eqnarray}
Z_\sigma = Z_g \, .
\end{eqnarray}
Renormalization of Eq.~(\ref{hPrime}) yields
\begin{align}
\label{hPrime2}
h^{\prime}  &  =h- (2gA + Z_g)\sigma c  - (2gA + Z_g) \frac{g}{2}c^{2} - (Bg+Z_\tau) \tau c 
\nonumber\\
&- (Cg +Z)\nabla^{2}c - (Dg -Z_w)w \nabla^{2}_\theta c \, .
\end{align}
Demanding that the renormalized $h^{\prime}$ retains its original from, cf. Eq.~(\ref{hPrime}), we obtain
\begin{subequations}
\begin{align}
A&=\frac{1-Z_{g}}{2g}\,,\quad B=\frac{1-Z_{\tau}}{g}\,,
\\
 C&=\frac{Z-1}{g}\,,\quad D=\frac{Z_w-1}{g} 
\end{align}
\end{subequations}
for the additive renormalization constants. Now we know all the renormalization factors and constants and hence we can write down the renormalized version of the Hamiltonian $\mathcal{H}_\sigma$, viz.\
\begin{eqnarray}
\mathcal{H}_\sigma = \mathcal{H} + (\sigma, \mathcal{A}) + O \left(  \sigma^2 \right)\, ,
\end{eqnarray}
where 
\begin{align}
\label{Adef}
\mathcal{A} &= \frac{Z_g}{2} \left( s^{2}- \frac{1}{N}\int_{\vec{\theta}}s^{2} \right) + \frac{Z_{\tau}-1}{g} \, \tau s 
\nonumber \\
&- \frac{Z-1}{g} \, \nabla^2 s - \frac{Z_w-1}{g} \, w \nabla^2_\theta s 
\end{align}
is the fully renormalized version of the operator $\mathcal{D}_1$. Naively, one might have expected that a mere multiplication with $Z^{-1}$ were sufficient to renormalize $\mathcal{D}_1$.  Equation~(\ref{Adef}), however, shows clearly that this is not the case.

Next we determine the scaling dimension of $\mathcal{A}$ via the scaling behavior of its coupling $\sigma$. Upon taking the derivative of Eq.~(\ref{sigmaRen}) with respect to the external inverse length scale $\mu$ we obtain
\begin{eqnarray}
\hat{\kappa} = \mu \partial_\mu \sigma |_0 = \gamma - \gamma_g \, .
\end{eqnarray}
Recalling that $\gamma^\ast = \eta$ and deducing from
\begin{eqnarray}
\beta = ( -\varepsilon +3 \gamma - 2 \gamma_g ) u 
\end{eqnarray}
that $\gamma_g^\ast = - \varepsilon/2 + 3 \eta /2$ we find 
\begin{eqnarray}
\hat{\kappa}^\ast = \frac{\varepsilon-\eta}{2} 
\end{eqnarray}
as the fixed point value of $\hat{\kappa}$. Taking into account that the naive dimension of $\sigma$ is two, we get that the RG flow of $\sigma$ in the vicinity of the fixed point is given by
\begin{eqnarray}
\sigma (\ell) = \sigma \, \ell^{-y_\sigma}
\end{eqnarray}
with
\begin{eqnarray}
y_\sigma = \frac{d-2+\eta}{2} \, .
\end{eqnarray}
Finally, this leads to
\begin{eqnarray}
x_{\mathcal{A}} = d - y_\sigma = \frac{d+2-\eta}{2} 
\end{eqnarray}
for the scaling dimension of $\mathcal{A}$. Note that $x_{\mathcal{A}}$ is identical to the scaling dimension of the operator $\mathcal{B}_3$.

We have to point out that $\mathcal{A}$ is not an eigenoperator of the RG. This fact follows from two observations: (i) $\mathcal{B}_3 = \mathcal{H}^\prime$ is an eigenoperator, as can be seen from the Ward identity~(\ref{wardHPrime}), and (ii) $\mathcal{A}$ is just one ingredient of $\mathcal{B}_3$,
\begin{eqnarray}
\mathcal{B}_3 = g \mathcal{A} + \tau s - \nabla^2 s - w \nabla_\theta^2 s \, ,
\end{eqnarray}
where, of course, all quantities in this equation are renormalized quantities. For completeness we mention that the Ward identity~(\ref{wardHPrime}) takes on the form
\begin{equation}
\label{wardA}
g\, \Gamma^{(n)}_{\mathcal{A}(\brm{p})} (\{ \brm{q} \})=-\Gamma^{(n+1)}  (\{ \brm{q} \}, \brm{p}) + \big( \brm{p}^2 + w \vec{\lambda}^2 + \tau  \big)\, \delta_{n,1}\, .
\end{equation}
when expressed in terms of $\mathcal{A}$. 

As announced above, we now briefly return to the simpler four-dimensional operators. The scaling dimension of $\mathcal{B}_0$ follows immediately from the scaling behavior of its coupling constant $\tau (\ell) = \tau \ell^{-y_\tau}$ with $y_\tau = 1/\nu$,
\begin{equation}
\label{scaleDimB0}
x_{\mathcal{B}_0} = d - y_\tau = d - 1/\nu \, .
\end{equation}
The operators $\mathcal{B}_1$ and $\mathcal{B}_2$ are obtained by applying, respectively, $w \nabla_\theta^2$ and $\nabla^2$ to the fundamental field $s$ and hence
\begin{subequations}
\begin{align}
x_{\mathcal{B}_1} &= x_s + 2 = \frac{d+2+\eta}{2} \, ,
\\
x_{\mathcal{B}_2} &= x_s + \frac{\phi}{\nu}= \frac{d-2+\eta}{2} + \frac{\phi}{\nu}  \, .
\end{align}
\end{subequations}
By now, we have expressed the scaling dimensions of all the four-dimensional operators which enter our analysis in Sec.~\ref{eom} in terms of the fundamental exponents $\eta$, $\nu$, and $\phi$. 

\subsection{Six-dimensional operators}
The most self-evident operators with naive dimension six are, perhaps,
\begin{align}
\mathcal{C}_2& = w s \nabla^2_\theta s \, , \quad \mathcal{C}_3 =  s \nabla^2 s \, , \quad \mathcal{C}_4 =  s^3 \, .
\end{align}
One eigen-combination of these operators follows immediately from Eq.~(\ref{anIdentity}), viz.\ $s \cdot \mathcal{H}^\prime$. Due to Eq.~(\ref{scalingRel}) we know that the scaling dimension of this eigenoperator is
\begin{equation}
x_{s\cdot\mathcal{H}^\prime} = d - x_s + x_s = d  \, ,
\end{equation}
i.e., it is marginal in any dimension. The physical reason for this marginality is that $s \cdot \mathcal{H}^\prime$ can be removed by a rescaling of the amplitude of the order parameter field.

Next, we take a closer look at the scaling dimension of the renormalized version of $\mathcal{D}_4$. Setting all the $f_i$ in Eq.~(\ref{Hamilt}) to zero, applying the renormalization scheme~(\ref{reno}) and taking the derivative with respect to the renormalized coupling constant $g$ we obtain
\begin{equation}
\label{derivH}
\frac{\partial \mathcal{H}}{\partial g} = \int d^d x \int_{\vec{\theta}} \mathcal{B}   \, ,
\end{equation}
where
\begin{align}
\label{defB}
\mathcal{B}&=\frac{1}{6}\Big(Z_{g}+2uZ_{g}^{\prime}\Big)s^{3} +\frac{uZ_{\tau}^{\prime}}{g}\tau s
\nonumber \\
&+ \frac{uZ^{\prime}}{g}(\nabla s)^{2} + \frac{uZ_w^{\prime}}{g}(\nabla_\theta s)^{2}
\end{align}
is the fully renormalized version $\mathcal{D}_4$. Here and in the following the prime indicates derivatives with respect to $u$. Equation~(\ref{derivH}) implies the Ward identity
\begin{equation}
\frac{\partial \Gamma^{(n)}}{\partial g} = - \Gamma^{(n)}_{\mathcal{B}}   \, .
\end{equation}
Taking into account that the genuine independent variables of the RGE for the vertex functions $\Gamma^{(n)}$ are $\mu$, $u$, $\tau$, and $w$, and exploiting the commutator
\begin{equation}
\left[  \frac{\partial}{\partial g},\mu\partial_{\mu}\right]  =\left[
G_{\varepsilon}^{1/2}\mu^{-\varepsilon/2}2\sqrt{u}\frac{\partial}{\partial
u},\mu\partial_{\mu}\right]  =\frac{\varepsilon}{2}\frac{\partial}{\partial g} \, ,
\end{equation} 
we find that the RGE for the vertex functions with an insertion of $\mathcal{B}$ is given by
\begin{align}
\label{RGEB}
&\Big(\mu\partial_{\mu}+\kappa\tau\partial_{\tau}+ \zeta w \partial_w +\beta\partial_{u}-\frac{n}
{2}\gamma+\beta^{\prime}+\frac{\varepsilon}{2}\Big)\Gamma^{(n)}_{\mathcal{B}
}
\nonumber \\
&=\frac{u}{g}\bigl(n\gamma^{\prime}-2\kappa^{\prime}\tau\partial_{\tau} -  2 \zeta^\prime w \partial_w\bigr)\Gamma^{(n)}\,.
\end{align}
Note that this RGE is not homogeneous and hence $\mathcal{B}$ is not an eigenoperator. Nevertheless, we can deduce the scaling dimension of $\mathcal{B}$ from the homogeneous part of the RGE~(\ref{RGEB}). Taking into account the operator's naive dimension $3(d-2)/2$, we obtain
\begin{equation}
x_{\mathcal{B}} = d + \omega  \, ,
\end{equation}
where $\omega$ is the Wegner exponent featured in Sec.~\ref{leader}.

\section{Calculation of Feynman diagrams}
\label{app:XXX}
In this appendix we give some details on the diagrammatic calculation that leads to Eqs.~(\ref{resGamma2}) and (\ref{resGamma3}). Instead of elaborating on all of the diagrams we restrict ourself to a couple of representative examples. The steps explained at these instances can then easily be adapted for the remaining diagrams.

\begin{widetext}
We start by considering diagram A with an insertion of ${\cal A}_1$, cf.\ Fig.~\ref{fig1}. Upon writing the conducting propagators in Schwinger representation we have
\begin{eqnarray}
\label{a1}
\mbox{A}_{{\cal A}_1} &=& g^2 w^2 \int^\infty_0 ds_1 ds_2 ds_3 \, \exp \left[ - (s_1 + s_2 + s_3) \tau \right] 
\nonumber \\
&\times&\int_{\brm{k}} \sum_{\vec{\kappa}} \left( \vec{\kappa}^2 \right)^2 \exp \big\{  - s_1 [\brm{k}^2 + w \vec{\kappa}^2] - s_2 [(\brm{k} + \brm{p})^2 + w \vec{\kappa}^2] - s_3 [(\brm{k} - \brm{q})^2 + w (\vec{\kappa} - \vec{\lambda})^2]\big\} \, .
\end{eqnarray}
Note that symmetry factor of $\mbox{A}_{{\cal A}_1}$ is 1 and not 1/2 (the symmetry factor of A), because there are 2 possibilities (the 2 conducting propagators) to insert ${\cal A}_1$. Now we carry out a completion of squares for the momenta as well as for the currents. After the straightforward momentum integration we arrive at
\begin{eqnarray}
\label{a2}
\mbox{A}_{{\cal A}_1} &=& g^2 w^2 \, \frac{1}{(4\pi)^{d/2}} \, \int^\infty_0 \frac{ds_1 ds_2 ds_3}{(s_1 + s_2 + s_3)^{d/2}} \, \exp \left[ - (s_1 + s_2 + s_3) \tau \right] \exp \left[ - \frac{s_2 s_3 (\brm{p} + \brm{q})^2 + s_1 s_2 \brm{p}^2 +  s_1 s_3 \brm{q}^2}{s_1 + s_2 + s_3} \right] 
\nonumber \\
&\times& \exp \left[ - R (s_1 , s_2 , s_3) w \vec{\lambda}^2 \right]  \sum_{\vec{\kappa}} \left[ \vec{\kappa} + \frac{s_3}{s_1 + s_2 + s_3} \vec{\lambda}\right]^4 \exp \left[ - (s_1 + s_2 + s_3) \vec{\kappa}^2 \right] \, , 
\end{eqnarray}
where $R (s_1 , s_2 , s_3) = (s_1 s_2 + s_2 s_3)/(s_1 + s_2 + s_3)$ is, according to our real-world interpretation, the total resistance of the diagram $\mbox{A}_{{\cal A}_1}$. At this stage it is useful to switch to continuous loop currents and to replace the summation $\sum_{\vec{\kappa}}$ by the integration $\int_{\vec{\kappa}}$. By standard Gaussian integration we then find for $D \to 0$ 
\begin{eqnarray}
\label{a3}
\mbox{A}_{{\cal A}_1} &=& g^2 w^2 \, \frac{1}{(4\pi)^{d/2}} \, \int^\infty_0 \frac{ds_1 ds_2 ds_3}{(s_1 + s_2 + s_3)^{d/2}} \, \exp \left[ - (s_1 + s_2 + s_3) \tau \right] \bigg\{ \frac{s_3^4}{(s_1 + s_2 + s_3)^4} \left(  \vec{\lambda}^2 \right)^2 + 2 \, \frac{s_3^2}{(s_1 + s_2 + s_3)^3} \frac{\vec{\lambda}^2}{w}
\nonumber \\
&-&  2 \, \frac{s_3^2}{(s_1 + s_2 + s_3)^4} \frac{\vec{\lambda}^2}{w} [s_2 s_3 (\brm{p} + \brm{q})^2 + s_1 s_2 \brm{p}^2 +  s_1 s_3 \brm{q}^2 + (s_1 s_3 + s_2 s_3) w \vec{\lambda}^2]
 \bigg\} \, , 
\end{eqnarray}
where we have carried out a Taylor expansion of the second and third exponential function appearing in Eq.~(\ref{a2}) and where we have discarded all convergent terms. The remaining integrations over the Schwinger parameters can be simplified by setting $s_1 = t x$, $s_2 = t y$ and $s_3 = t (1 -x - y)$ and then integrating $t$ from $0$ to $\infty$, $y$ from $0$ to $1-x$ and $x$ from $0$ to $1$. Expanding the so obtained intermediate result for small $\varepsilon$, we obtain  
\begin{eqnarray}
\label{a4}
\mbox{A}_{{\cal A}_1} = - g^2  w \vec{\lambda}^2\,  \frac{G\varepsilon}{\varepsilon} \, \tau^{-\varepsilon /2} \bigg\{  \frac{\tau}{3}  + \frac{\brm{p}^2}{90}  + \frac{\brm{p}_1^2 + \brm{p}_2^2}{30} \bigg\} \, , 
\end{eqnarray}
where we have set $\brm{p} = \brm{p}_1 +\brm{p}_2$ and $\brm{q} = -\brm{p}_2$. The computation of $\mbox{B}_{{\cal A}_1}$ is, in comparison, simple because it does not involve a summation over a loop current. The total 1-loop contribution to $-\Gamma _{{\cal A}^\prime_1 }^{(2)}$ is given by
\begin{eqnarray}
-\Gamma_{{\cal A}^\prime_1 }^{(2) \, \text{1-loop}} =  \mbox{A}_{{\cal A}_1} - 2 \, \mbox{B}_{{\cal A}_1} = - g^2  w \vec{\lambda}^2\,  \frac{G\varepsilon}{\varepsilon} \, \tau^{-\varepsilon /2} \bigg\{  \frac{\tau}{3}  + w \vec{\lambda}^2 + \frac{\brm{p}^2}{90}  + \frac{\brm{p}_1^2 + \brm{p}_2^2}{30} \bigg\} \, . 
\end{eqnarray}
Here we have a clear example of the mixing of irrelevant operators in our perturbation calculation. The insertion of ${\cal A}_1$ does not only generate primitive divergences proportional to $w^2 (\vec{\lambda}^2)^2$ but also those of the type $\tau \, w \vec{\lambda}^2$ and $\brm{p}^2 \, w \vec{\lambda}^2$. Note the we dropped the first term in the braces in Eq.~(\ref{resGamma2}) because it leads only to subdominant correction due to the factor $\tau$.

As a next example, we consider the diagram $\mbox{A}_{{\cal A}_2}$ that stands for
\begin{eqnarray}
\mbox{A}_{{\cal A}_2} &=& g^2 w \int^\infty_0 ds_1 ds_2 ds_3 \, \exp \left[ - (s_1 + s_2 + s_3) \tau \right] \int_{\brm{k}} \sum_{\vec{\kappa}}  \, \vec{\kappa}^2 \, \frac{\brm{k}^2 + (\brm{k} + \brm{p})^2}{2}
\nonumber \\
&\times& \exp \big\{  - s_1 [\brm{k}^2 + w \vec{\kappa}^2] - s_2 [(\brm{k} + \brm{p})^2 + w \vec{\kappa}^2] - s_3 [(\brm{k} - \brm{q})^2 + w (\vec{\kappa} - \vec{\lambda})^2]\big\} \, .
\end{eqnarray}
Completion of squares in the momenta and the currents leads to 
\begin{eqnarray}
\mbox{A}_{{\cal A}_2} &=& \frac{g^2}{2} \, w  \int^\infty_0 ds_1 ds_2 ds_3 \exp \left[ - (s_1 + s_2 + s_3) \tau \right] \exp \left[ - \frac{s_2 s_3 (\brm{p} + \brm{q})^2 + s_1 s_2 \brm{p}^2 +  s_1 s_3 \brm{q}^2}{s_1 + s_2 + s_3} \right] 
\nonumber \\
&\times& \exp \left[ - R (s_1 , s_2 , s_3) w \vec{\lambda}^2 \right]  \int_{\vec{\kappa}} \left[ \vec{\kappa} + \frac{s_3}{s_1 + s_2 + s_3} \vec{\lambda}\right]^2 \exp \left[ - (s_1 + s_2 + s_3) \vec{\kappa}^2 \right] 
\nonumber \\
&\times& \int_{\brm{k}} \left[ \left( \brm{k} - \frac{s_2 \brm{p} - s_3 \brm{q}}{s_1 + s_2 + s_3} \right)^2 + \left( \brm{k} + \brm{p} - \frac{s_2 \brm{p} - s_3 \brm{q}}{s_1 + s_2 + s_3} \right)^2 \right] \exp \left[ - (s_1 + s_2 + s_3) \brm{k}^2 \right] \, , 
\end{eqnarray}
where we switched to continuous loop currents. Carrying out both Gaussian integrations and performing a Taylor expansion we obtain in the replica limit
\begin{eqnarray}
\mbox{A}_{{\cal A}_2} &=&  \frac{g^2}{2}\,  w \vec{\lambda}^2 \, \frac{1}{(4\pi)^{d/2}} \, \int^\infty_0 \frac{ds_1 ds_2 ds_3}{(s_1 + s_2 + s_3)^{d/2}} \, \exp \left[ - (s_1 + s_2 + s_3) \tau \right] \bigg\{ \frac{d \, s_3^2}{(s_1 + s_2 + s_3)^3}  + \frac{s_3^2 \, \brm{p}^2}{(s_1 + s_2 + s_3)^2} 
\nonumber \\
&+& 2 \, \frac{s_2^2 s_3^2 \brm{p}^2 - 2 s_2 s_3^3 \brm{p} \cdot \brm{q} + s_3^4 \brm{q}^2}{(s_1 + s_2 + s_3)^4} - 2 \, \frac{s_2 s_3^2 \brm{p}^2 -  s_3^3 \brm{p} \cdot \brm{q} }{(s_1 + s_2 + s_3)^3} - d \, \frac{s_2 s_3^3 (\brm{p} + \brm{q})^2 + s_1 s_2 s_3^2 \brm{p}^2 + s_1 s_3^3 \brm{q}^2 }{(s_1 + s_2 + s_3)^4} 
\nonumber \\
&-& d \, \frac{(s_1 s_3^3 + s_2 s_3^3)w \vec{\lambda}^2 }{(s_1 + s_2 + s_3)^4}
 \bigg\} \, 
\end{eqnarray}
where we have discarded convergent terms. Using the same change of variables as for $\mbox{A}_{{\cal A}_1}$ we integrate out the Schwinger parameters. This leads in $\varepsilon$ expansion to the result
\begin{eqnarray}
\mbox{A}_{{\cal A}_2}  = - g^2  w \vec{\lambda}^2\,  \frac{G\varepsilon}{\varepsilon} \, \tau^{-\varepsilon /2} \bigg\{  \frac{\tau}{2}  + \frac{w \vec{\lambda}^2}{10} - \frac{\brm{p}^2}{90}  + \frac{\brm{p}_1^2 + \brm{p}_2^2}{60} \bigg\} \, . 
\end{eqnarray}
The diagram $\mbox{B}_{{\cal A}_2}$ can be calculated by similar means. For the entire 1-loop contribution to $-\Gamma _{{\cal A}^\prime_2}^{(2)}$ we find
\begin{eqnarray}
-\Gamma_{{\cal A}^\prime_2}^{(2) \, \text{1-loop}} =  \mbox{A}_{{\cal A}_2} - 2 \, \mbox{B}_{{\cal A}_2} =  g^2  w \vec{\lambda}^2\,  \frac{G\varepsilon}{\varepsilon} \, \tau^{-\varepsilon /2} \bigg\{  \frac{5 \, \tau}{2}  + \frac{19 \, w \vec{\lambda}^2}{10} + \frac{\brm{p}^2}{90}  + \frac{3\, ( \brm{p}_1^2 + \brm{p}_2^2)}{20} \bigg\} \, . 
\end{eqnarray}

Diagram $\mbox{A}_{{\cal A}_3}$ serves as our final example. In Schwinger representation this diagram reads
\begin{eqnarray}
\mbox{A}_{{\cal A}_3} &=& - g \, w \int^\infty_0 ds_1 ds_2  \, \exp \left[ - (s_1 + s_2 ) \tau \right] \int_{\brm{k}} \sum_{\vec{\kappa}}  \, \frac{\vec{\kappa}^2 + (\vec{\kappa} - \vec{\lambda})^2 + \vec{\lambda}^2}{3}
\nonumber \\
&\times& \exp \big\{  - s_1[(\brm{k} - \brm{q})^2 + w (\vec{\kappa} - \vec{\lambda})^2]  - s_2 [(\brm{k} + \brm{p})^2 + w \vec{\kappa}^2] \big\} \, .
\end{eqnarray}
Once more we undertake a completion of squares in the momenta and currents. After integrating out the loop momentum we have
\begin{eqnarray}
\mbox{A}_{{\cal A}_3} &=& - \frac{g}{3} \,  w \, \frac{1}{(4 \pi)^{d/2}} \int^\infty_0 \frac{ds_1 ds_2}{(s_1 + s_2 )^{d/2}}  \, \exp \left[ - (s_1 + s_2 ) \tau \right] \exp \left[  - R (s_1, s_2) \, [w \vec{\lambda}^2 + (\brm{p} + \brm{q})^2] \right]
\nonumber \\
&\times&  \int_{\vec{\kappa}} \bigg[ \vec{\lambda}^2 + \Big( \vec{\kappa} - \frac{s_2}{s_1 + s_2} \vec{\lambda} \Big)^2 +  \Big( \vec{\kappa} + \frac{s_1}{s_1 + s_2} \vec{\lambda} \Big)^2 \bigg] \exp \left[ - (s_1 + s_2 ) w \vec{\kappa}^2 \right] 
 \, ,
\end{eqnarray}
with the total resistance of this diagram being $R (s_1, s_2) = s_1 s_2 / (s_1 + s_2 )$. Integration over the loop current and Taylor expansion gives, up to convergent terms,
\begin{eqnarray}
\mbox{A}_{{\cal A}_3} &=& - \frac{g}{3} \,  w \vec{\lambda}^2 \, \frac{1}{(4 \pi)^{d/2}} \int^\infty_0 \frac{ds_1 ds_2}{(s_1 + s_2 )^{d/2}}  \, \exp \left[ - (s_1 + s_2 ) \tau \right] \bigg\{ 1 + 2 \, \frac{s_1^2}{(s_1 + s_2 )^2} 
\nonumber \\
&- & R (s_1, s_2)\,  [w \vec{\lambda}^2 + (\brm{p} + \brm{q})^2]  -2\, \frac{s_1^3 s_2}{(s_1 + s_2 )^3}\,  [w \vec{\lambda}^2 + (\brm{p} + \brm{q})^2] \bigg\} \, .
\end{eqnarray}
The integrations over the Schwinger parameters can here to simplified by setting $s_1 = t x$ and $s_2 = t (1-x)$. After symmetrizing the external momenta the $\varepsilon$ expanded result reads 
\begin{eqnarray}
\mbox{A}_{{\cal A}_3}  =  g \,  w \vec{\lambda}^2\,  \frac{G\varepsilon}{\varepsilon} \, \tau^{-\varepsilon /2} \bigg\{  \frac{10 \, \tau}{9}  + \frac{8 \, w \vec{\lambda}^2}{45}  + \frac{8\, (\brm{p}_1^2 + \brm{p}_2^2)}{90} \bigg\} \, . 
\end{eqnarray}
The calculation of $\mbox{B}_{{\cal A}_3}$ is fairly easy. Merging the 2 results we obtain
\begin{eqnarray}
-\Gamma_{{\cal A}^\prime_3}^{(2) \, \text{1-loop}} =  \mbox{A}_{{\cal A}_3} - 2 \, \mbox{B}_{{\cal A}_3} =  - g  \, w \vec{\lambda}^2\,  \frac{G\varepsilon}{\varepsilon} \, \tau^{-\varepsilon /2} \bigg\{  \frac{14 \, \tau}{9}  + \frac{52 \, w \vec{\lambda}^2}{45}  + \frac{2 \, ( \brm{p}_1^2 + \brm{p}_2^2)}{15} \bigg\} \, . 
\end{eqnarray}

The 3-leg diagram with insertions can be computed by similar techniques as we have used for the 2-leg diagrams. The 3-leg diagram benefit from the extra simplification that they can be evaluated at vanishing external momenta because all their contributions with non zero external momenta are convergent. 
\end{widetext}


\begin{thebibliography}{}
\bibitem{footnote1}
The $|_0$ indicates that bare quantities are kept fixed while taking the derivatives.
\bibitem{footnote2}
Strictly speaking, we derived the scaling behavior of the $n$-point correlation functions. To switch between vertex and correlation functions, however, one just needs to change the sign of the exponent $(d-2+\eta)n/2$.
\end{thebibliography}
\end{document}